\documentclass[a4,prb,onecolumn,superscriptaddress,amsfonts,amssymb]{revtex4-1}
\usepackage{graphicx}

\newcommand{\mv}{B} 

\begin{document}


\title{Eckhaus-like instability of large scale coherent structures \\ 
in a fully turbulent von K\'arm\'an flow} 



\author{E.~Herbert}
\email[]{eric.herbert@univ-paris-diderot.fr} \affiliation{SPHYNX, Service de
Physique de l'\'Etat Condens\'e, DSM, CEA Saclay, CNRS URA 2464,
91191 Gif-sur-Yvette, France}
\author{P.-P.~Cortet}
\affiliation{Laboratoire FAST, Universit\'{e} Paris-Sud, CNRS,
France}
\author{F.~Daviaud}
\affiliation{SPHYNX, Service de Physique de l'\'Etat Condens\'e,
DSM, CEA Saclay, CNRS URA 2464, 91191 Gif-sur-Yvette, France}
\author{B.~Dubrulle}
\affiliation{SPHYNX, Service de Physique de l'\'Etat Condens\'e, DSM, CEA
Saclay, CNRS URA 2464, 91191 Gif-sur-Yvette, France}


\date{\today}

\begin{abstract}
The notion of instability of a turbulent flow is introduced in the
case of a von K\'arm\'an flow thanks to the monitoring of the
spatio-temporal spectrum of the velocity fluctuations, combined
with projection onto suitable Beltrami modes. It is shown that the
large scale coherent fluctuations of the flow obeys a
sequence of Eckhaus instabilities when the Reynolds number
$\mathrm{Re}$ is varied from $10^2$ to $10^6$. This sequence
results in modulations of increasing azimuthal wavenumber. The
basic state is the laminar or time-averaged flow at an
arbitrary $\mathrm{Re}$, which is axi-symmetric, {\it i.e.} with a
$0$ azimuthal wavenumber. Increasing $\mathrm{Re}$ leads to
non-axisymmetric modulations with increasing azimuthal wavenumber
from $1$ to $3$. These modulations are found to rotate in the
azimuthal direction. However no clear rotation frequency can be
established until $\mathrm{Re}\approx 4\times 10^3$. Above, they
become periodic with an increasing frequency. We finally show that
these modulations are connected with the coherent structures of
the mixing shear layer. The implication of these findings for the
turbulence parametrization is discussed. Especially, they may
explain why simple eddy viscosity models are able to capture
complex turbulent flow dynamics.
\end{abstract}


\maketitle 

\section{Introduction}
In classical phenomenology, turbulence arises after a sequence of
symmetry breakings, successive instabilities or bifurcations,
which however progressively restore the system symmetries in a
statistical sense.~\cite{frisch_turbulence_1995} The study of
these instabilities traditionally proceeds from (linear or
non-linear) perturbations of the so-called ``basic state", the
stationary laminar solution of the Navier-Stokes equation at low
Reynolds number. At finite Reynolds number there is no general
well-defined criterion to discriminate between the turbulent or
laminar nature of the flow: one cannot define clearly a critical
threshold beyond which the flow is turbulent and below which it is
laminar. The consensus view is that the flow is turbulent when the
Reynolds number is large enough and when a well established
spatio-temporal energy spectrum is observed with broad-band power
laws. In such a case a possible statistical equivalent of the
laminar ``basic state'' can be defined using the (statistically or
time) averaged flow. This flow is stationary by construction, but
differs from an usual basic state in the sense that it is solution
of the ensemble averaged Navier-Stokes equation instead of a solution of the plain Navier-Stokes equation. A
natural question then arises: what happens once the statistically
stationary turbulent state is reached? Is this the end of the
story or can new instabilities of the  averaged turbulent flow
develop? Experimentally, the answer seems to be positive, since
spontaneous bifurcations and flow reversals in fully developed
turbulence have already been observed: i) in a wake flow,
there is a mean  pattern transition at a critical Reynolds
number. This corresponds to the so-called ``drag-crisis'' (see
Ref.~\onlinecite{tritton_physical_1977}). This leads to a dramatic
decrease of the mean drag of a sphere or a cylinder in a turbulent
flow for a critical value of the Reynolds number
$\mathrm{Re}\sim10^5$. The wake becomes narrower as the mean flow
pattern changes. ii) Spontaneous flow reversals have been observed
in thermally driven (Rayleigh-B\'enard) convection, both
experimentally\cite{sugiyama_flow_2010} and
numerically.\cite{poel_flow_2012,podvin_proper_2012} In a
cylindrical Rayleigh-B\'enard geometry the reversals of Large
Scale Circulation was reported to be induced by reorientation
along the azimuthal direction.\cite{mishra_dynamics_2011} However
it is difficult to experimentally characterize a bifurcation from
a non-stationary or periodic flow. iii) In a magnetohydrodynamic
turbulence, the large scale magnetic field spontaneously
generated in a liquid metal at $\mathrm{Re}>10^6$ was
shown\cite{berhanu_magnetic_2007} to undergo spontaneous
reversals, with a dynamics  governed by a few magnetic modes
despite the strongly turbulent background. This observation led to
a simple model of geodynamos, with a few modes undergoing chaotic
dynamics.\cite{gallet_reversals_2012} An open problem common to
all these situations is to find suitable tools to study and
characterize these type of instabilities.

The purpose of this paper is to present some answers to these
questions in the specific case of a von K\'arm\'an flow. This flow
is generated by two counter-rotating impellers in a cylindrical
vessel. At low Reynolds number, the laminar flow is axi-symmetric
and divided into two toric recirculation cells separated by an
azimuthal shear layer. Its transition to turbulence for various
counter-rotations and aspect ratio has been extensively studied
theoretically, numerically and experimentally, with smooth or
rough (fitted with blades) impellers (see e.g.
Refs.~\onlinecite{batchelor_note_1951,stewartson_flow_1953,zandbergen_vonkarman_1987,nore_12_2003,nore_survey_2004,
moisy_experimental_2004,lopez_instability_2002,ravelet_supercritical_2008}
and references therein). The study in
Ref.~\onlinecite{ravelet_supercritical_2008} was performed in the
exact-counter-rotating case. It was reported that a first
bifurcation from the stationary and axi-symmetric laminar state
occurs at $\mathrm{Re}=175$, yielding a stationary flow with an
azimuthal modulation, \textit{i.e.} for which the axisymmetry is
broken. This stationary state persists up to typically
$\mathrm{Re} \sim 300$ where time-dependence arises. The
transition to turbulence further proceeds through plain,
modulated or chaotic traveling waves, until $\mathrm{Re}\sim
10^4$, where a ``fully developed turbulent state" seems to have
been reached. In that state, the energy spectrum
is broad and the dimensionless dissipation does not depend on
Reynolds number anymore.~\cite{ravelet_supercritical_2008} In that turbulent state, the
flow once time averaged regains the structure of the basic laminar
flow, made of two shearing toric recirculation cells. In the
present paper, we have worked using the same experimental setup.
We demonstrate that the flow is organized into large scale
coherent structures. These structures are subject to a sequence
of Eckhaus type instabilities leading to an increasing azimuthal
wavenumber as the Reynolds number varies from $10^2$ to $10^6$,
similar to the sequence of instabilities at much lower
$\mathrm{Re}$. Since this sequence of instabilities occurs on
coherent structures in a turbulent flow rather than a laminar
``basic state", we are facing a new paradigm that both requires
new tools of investigations and opens interesting questions about
turbulence parametrization. The tools are presented in
Section~\ref{sec:technical}, the instability is described in
Section~\ref{sec:results} and some theoretical consequences are
discussed in Section~\ref{sec:theocons}.


\section{Technical background and tools}
\label{sec:technical}
\subsection{Experimental setup}
We have worked with a von K\'arm\'an flow generated by two coaxial
and counter-rotating impellers in a cylindrical vessel. More
details about the experimental setup can be found in
Ref.~\onlinecite{cortet_susceptibility_2011}. The cylinder radius
and height are $R=100$\,mm and $H=180$\,mm respectively. The
impellers consist of $185$\,mm diameter disks fitted with sixteen
curved blades of $20$\,mm height. The impellers are driven by two
independent motors. The motors frequencies are respectively set to
$f_1$ and $f_2$.
In the present work the impellers are rotating with the convex
face of the blades going forwards, contrarily to
Ref.~\onlinecite{ravelet_supercritical_2008}.
The working fluid is either pure water, a water-glycerol mixture
(26\%-74\% in weight respectively) or pure glycerol. The resulting
accessible Reynolds numbers, $\mathrm{Re}=\pi (f_1+f_2) R^2 \nu^{-1}$
with $\nu$ the kinematic viscosity, vary from $10^2$ to $10^6$.

\begin{figure}[ht]
\centerline{\includegraphics[width=0.35\textwidth]{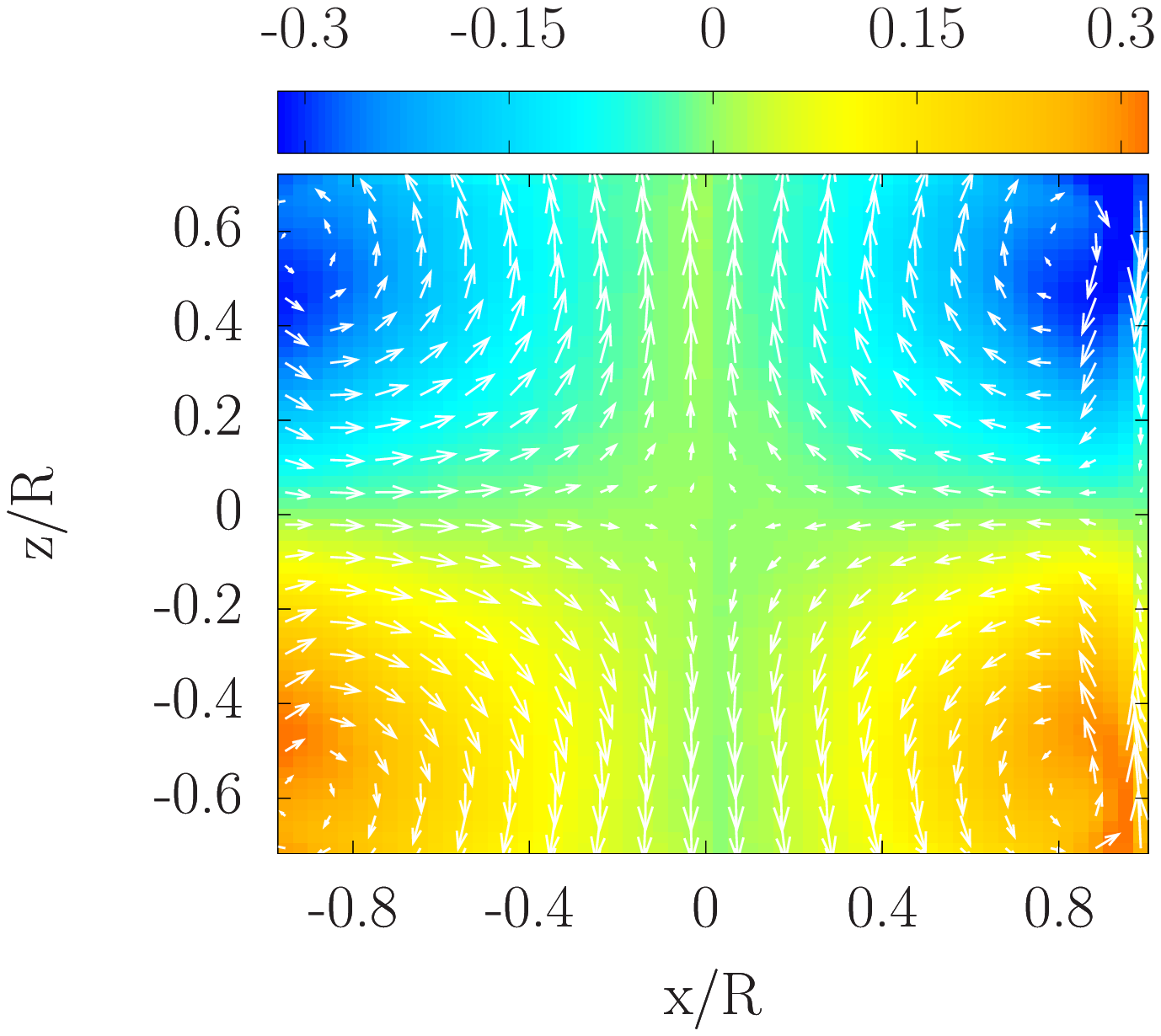} 
\hspace{1cm} 
\includegraphics[width=0.35\textwidth]{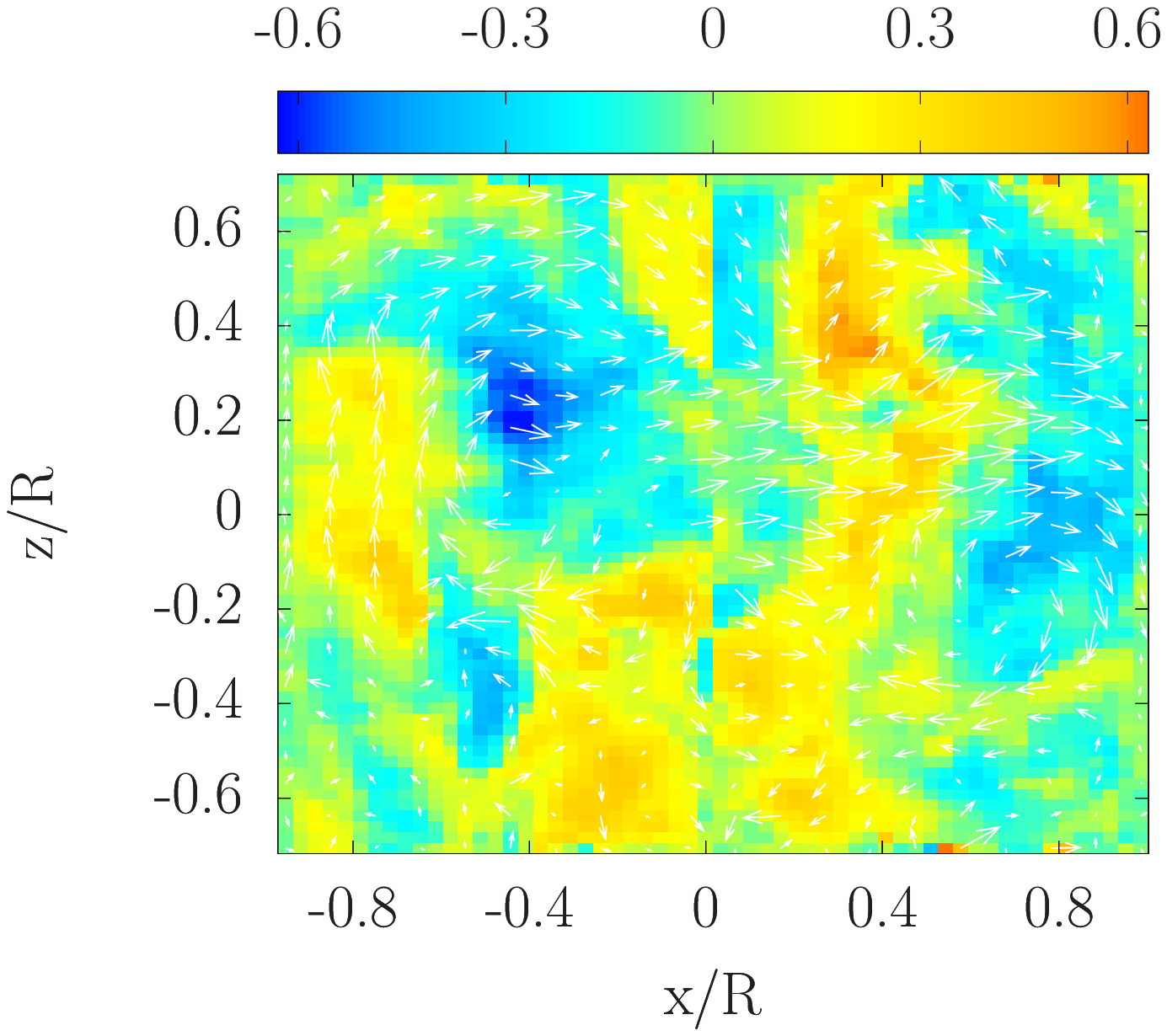}}
     \caption{(Color online) Left, time-averaged velocity field ${\bf
     \bar{v}}$, measured in a meridian plane using S-PIV, at
     $\mathrm{Re}=10^6$ and $\theta=(f_1-f_2)/(f_1+f_2)=0$. $x=r\cos \varphi_0$,
     $\varphi_0=[0,\pi]$. Right, corresponding instantaneous
     fluctuating component of the velocity field ${\bf v'}={\bf v}-{\bf
     \bar{v}}$. The color maps the azimuthal velocity $v_\varphi$
     (normalized by $R(f_1+f_2)/2$) whereas the arrows map the
     ($v_x$,$v_z$) velocities.}\label{fig:velocityfield_exp}
\end{figure}

This setup is invariant under $\cal R_{\pi}$ rotations around any
radial axis passing through the center of the cylinder.
Measurements are done thanks to a Stereoscopic Particle Image
Velocimetry (S-PIV) system. The S-PIV provides time series of the
3 components of the velocity (radial $v_x(x,z,t)$, vertical
$v_z(x,z,t)$ and azimuthal $v_y(x,z,t)$)  on a $63\times 58$
points grid in a meridian plane $[x,z]$ composed of two planes
dephased by $\pi$ in the cylindrical coordinates. $x=r \cos
(\varphi_0)$, with the azimuthal angle $\varphi_0=[0,\pi]$ and $r$
the radial distance from the impellers axis. $z$ is the vertical
distance from the center of the cylinder, $y$ is the out-of-plane
direction and $t$ is the time. The azimuthal component is then
written in the cylindrical coordinates using $v_\varphi(x,z,t) =
x/|x|\,v_y(x,z,t)$. The size of the recorded images is $194.9
\times 144.3$\,mm$^2=58\,\delta_R \times 63\,\delta_Z$
(where $\delta_R=3.3$\,mm and $\delta_Z=2.3$\,mm are the
spatial resolutions of the S-PIV). Time series have sampling
frequency in the range $f_s=1.7$ to 15\,Hz and are composed of
$N=1200$ to 4200 samples. Each experiment is started with
the impellers and the fluid at rest. The impellers velocities are
then suddenly increased to reach their target values. After a few
tens of seconds, a statistically stationary state is reached and
S-PIV time series is finally acquired.


\subsection{Experimental perturbation analysis}
For low Reynolds number $\mathrm{Re}\le 500$, the flow is steady
and laminar and only few fluctuations are present.  When $\theta =
(f_1-f_2)/(f_1+f_2)=0$, the instantaneous flow is composed of two
toric recirculation cells separated by an azimuthal shear layer
located at $z=0$ reflecting the $R_\pi$-symmetry of the system.
Increasing the Reynolds number, one expects to reach fully
developed turbulence around $\mathrm{Re}=10^4$ as observed in
Ref.~\onlinecite{ravelet_supercritical_2008}, with increasing
difference between the instantaneous flow and the time-averaged
flow. Above $\mathrm{Re}=500$ and at $\theta=0$, the axisymmetry
is broken by fluctuations  of the instantaneous flow. However,
this symmetry is restored for the time-averaged flow (\emph{cf.}
Fig.\,\ref{fig:velocityfield_exp}), at any Reynolds number. By
construction, this average flow is time-independent and will be the equivalent of a ``basic
state". As the Reynolds number is increased, this ``basic state"
is subject to increasingly stronger
fluctuations.~\cite{monchaux_properties_2006} The question we want
to address here is: are these fluctuations purely disorganized as
in thermal noise or can we identify a pattern in them when varying
the Reynolds number,\cite{he_thermodynamical_1998} in a way
similar to a sequence of bifurcations of ordinary instabilities?
To answer this question we need two steps: i)  to separate the
fluctuations into ``organized" motions and ``thermal" noise; ii)
to build a suitable tool to study the temporal behavior of these
organized motions. These steps are described below. For step i),
we use comparison with and projection onto Beltrami modes. For
step ii), we use spatio-temporal spectra, as used in wave
turbulence.\cite{cobelli_space-time_2009}

\begin{figure}[h]
\centering
\begin{tabular}{cccc}
   \includegraphics[width=0.27\textwidth]{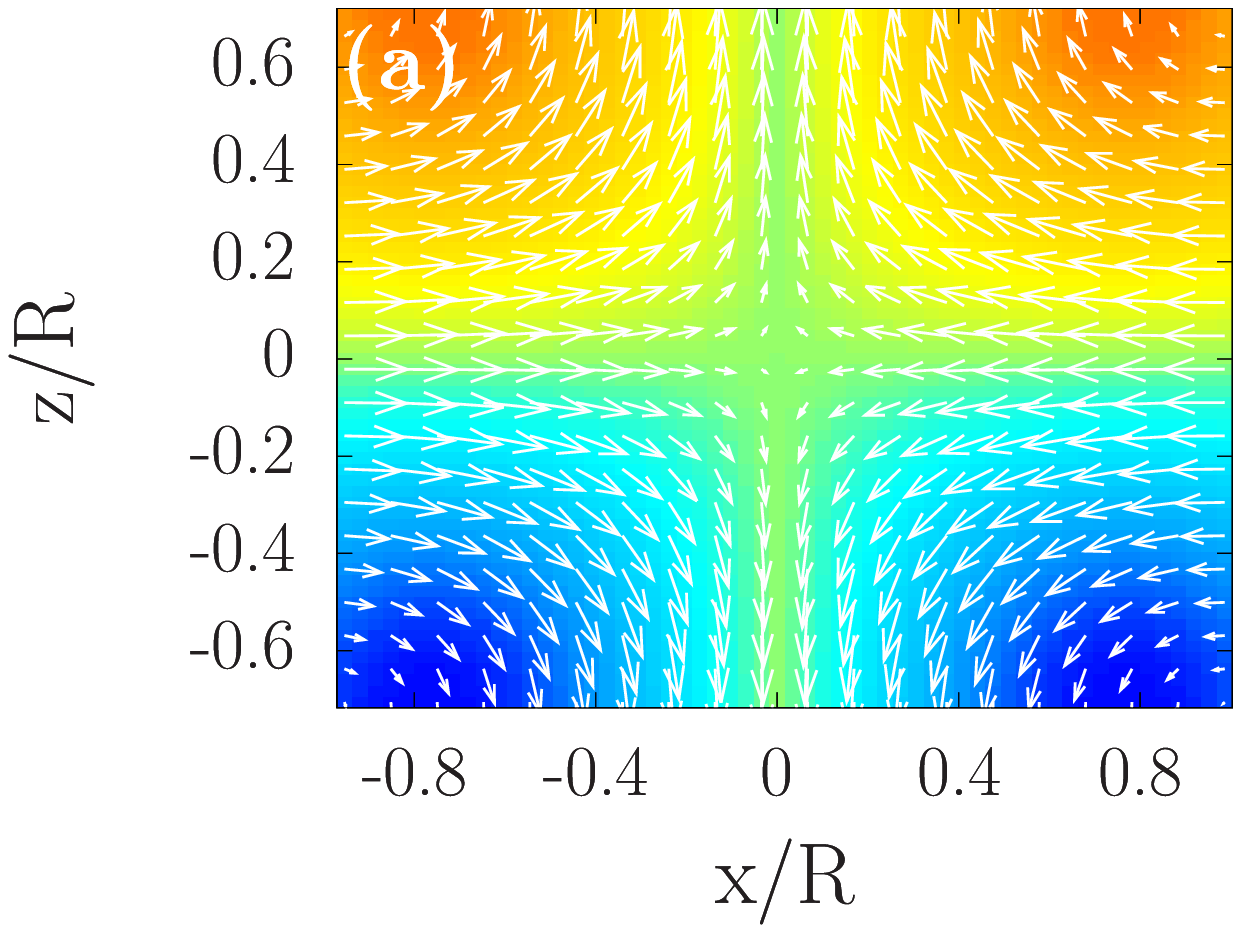}  &
   \includegraphics[width=0.23\textwidth]{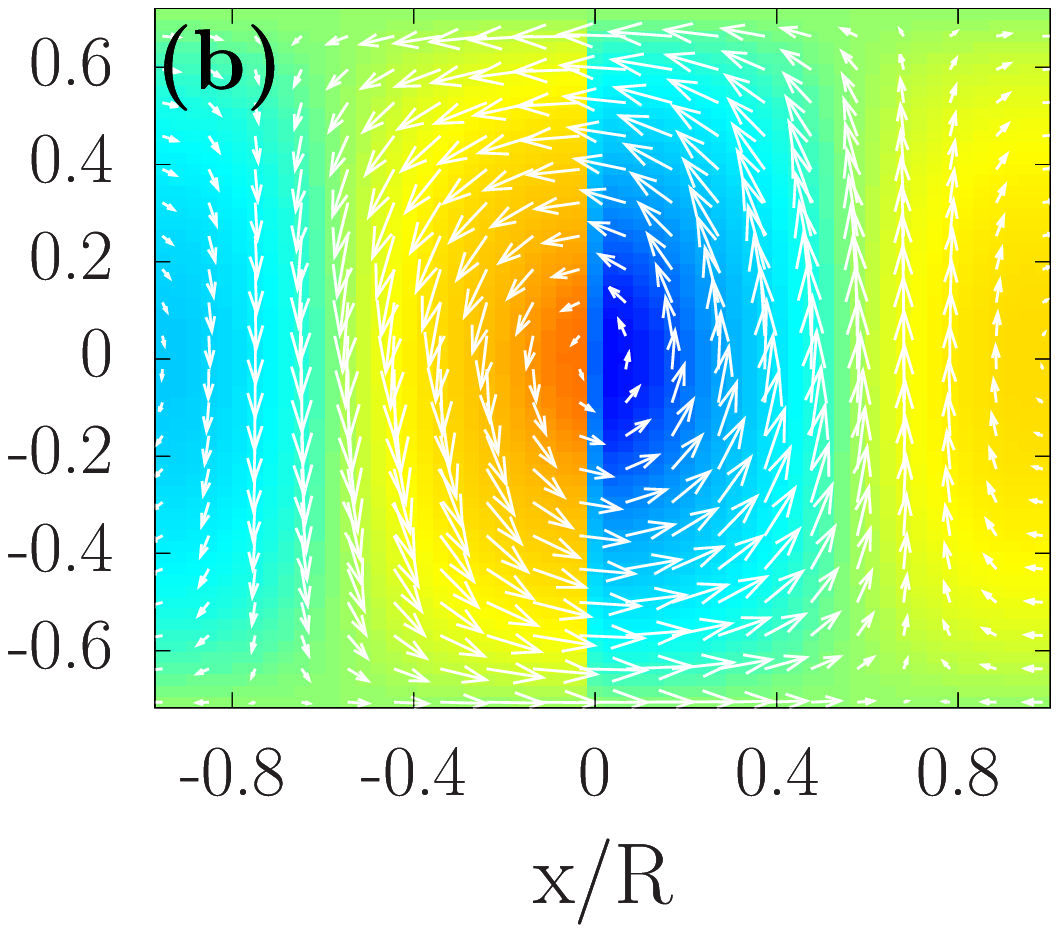}  &
   \includegraphics[width=0.23\textwidth]{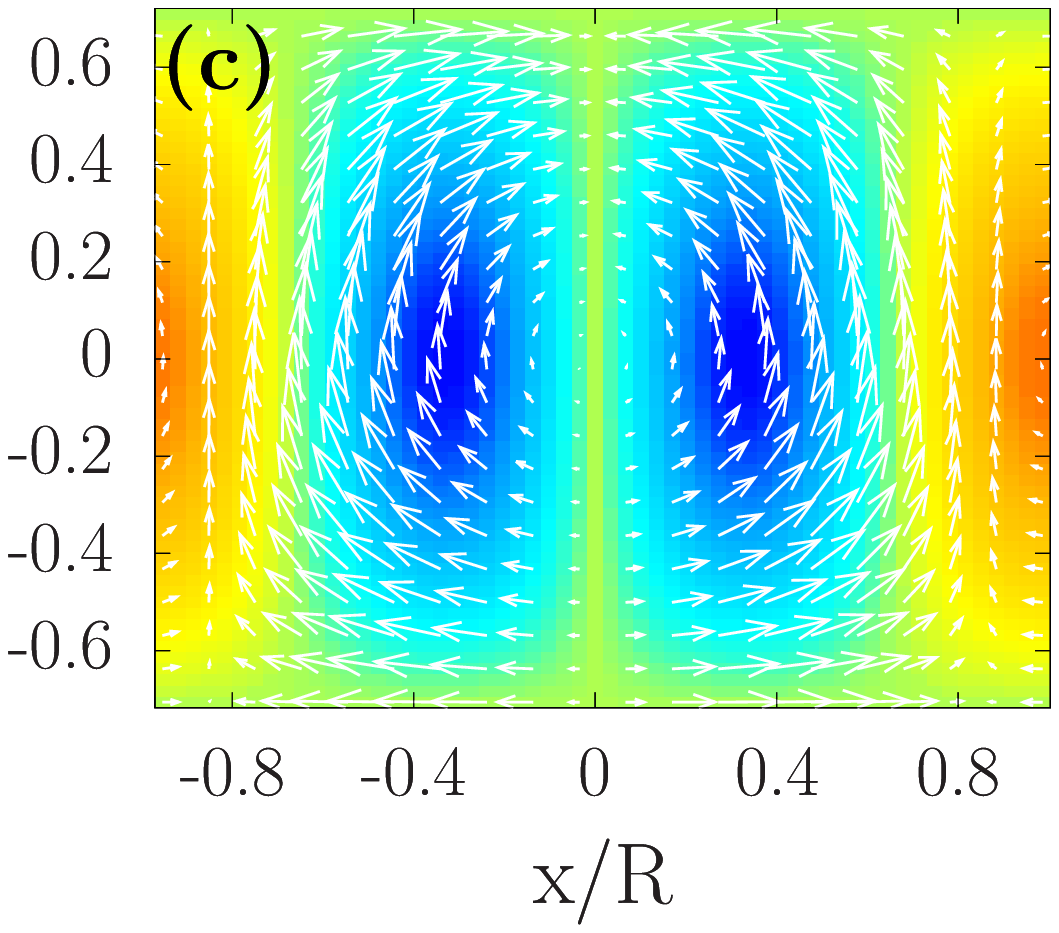}  &
   \includegraphics[width=0.23\textwidth]{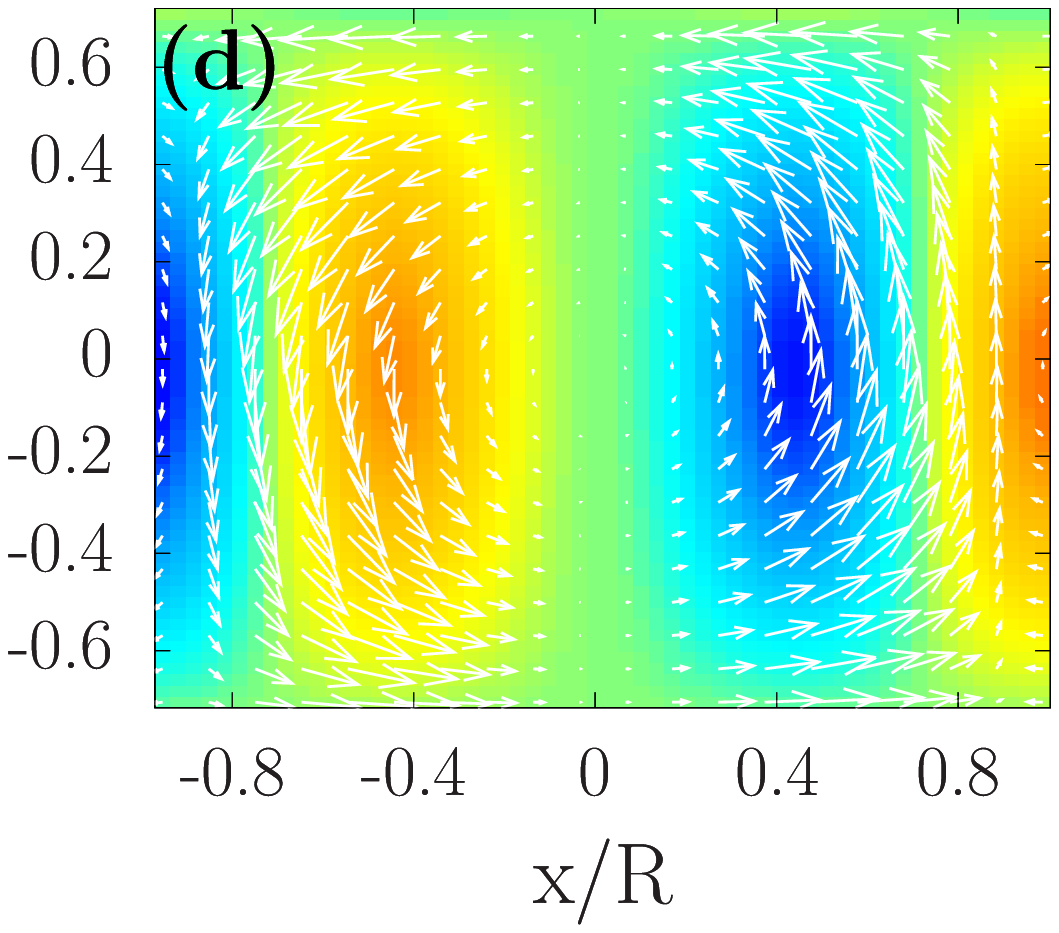}  \\
\end{tabular}
     \caption{     (Color online) (a) (resp. (b), (c) and (d)) synthetic
     velocity field ${\bf b^{nmk}}$ obtained from corresponding
     Beltrami modes ${\bf B^{nmk}}$ in a meridional plane with
     $n=k/k_0=1$ and $m=0$ (resp. $m=1$, $2$ and $3$). Same color and
     arrow codes as in Fig.\,\ref{fig:velocityfield_exp}.
     }\label{fig:velocityfield}
\end{figure}


\subsubsection{Projection onto Beltrami modes}

The measured velocity fields time series are analyzed thanks to a
projection on the basis of Beltrami modes as detailed in
Ref.\,\onlinecite{boisson_three-dimensional_2011}. Beltrami modes
have been introduced as a general spectral decomposition
basis,~\cite{Constantin_beltrami_1988} and any velocity field in a
cylindrical geometry can be decomposed as a superposition of such
modes. Given a vector field, ${\bf v}$, lying in a cylinder of
radius $R=1$ and height $H=2h$, with usual cylindrical components
${\bf v}=(v_r,v_\varphi,v_z)$, we first switch to the new
components ${\bf V}=(V_+,V_-,v_z)$ such that $V_\pm=(v_r\pm i
v_\varphi)/2$. The field
${\bf V}$ can further be decomposed over Beltrami modes ${\bf
B^{nmks}}$ as:
\begin{equation}
{\bf V} = \sum_{n=1}^N \sum_{m=-M}^M \sum_{k=-Pk_0}^{Pk_0} 
\sum^{(+)}_{s=(-)} D_{nmk} {\bf \mv^{nmks}},
\label{decompoBeltrami}
\end{equation}
where $k_0= \pi/h$, $n$ and $m$ are integer numbers and ($N,M,P$)
are the number of considered modes in the ($r,\varphi,z$)
directions respectively. The $D_{nmk}$ are complex
amplitudes which conjugates verifies $D_{nmk}^*=D_{n,-m,-k}$ and
the Beltrami modes are given by
\begin{equation}
{\bf \mv^{nmks}}=\left(
\begin{array}{c}
\mv_+^{nmks}\\
\mv_-^{nmks}\\
\mv_z^{nmk}\\
\end{array}\right)=
  \frac{1}{2}\left(
\begin{array}{c}
\left(\lambda^s -k\right)J_{m+1}(\mu_{nm} r)\\
\left(\lambda^s +k\right)J_{m-1}(\mu_{nm} r)\\
-2i\mu_{nm}J_{m}(\mu_{nm} r)\\
\end{array}\right)\exp(i m \varphi +i k z).
\label{BeltramiExpr}
\end{equation}
Here, $\lambda^{s}=s\sqrt{\mu_{nm}^2+k^2}$ with $s=\pm$ and $J_m$
is the Bessel function of order $m$. The orthogonality condition
is ensured by the fact the $\mu_{nm}$ coefficients are the $n^{\rm
th}$ root of $J_m$. Eq.~(\ref{decompoBeltrami}) then corresponds
to a decomposition into solenoidal Beltrami waves with
polarization given by the sign $s$ of $\lambda^s$. From now on, we will omit the $s$ superscript for simplicity.
 We
define the synthetic velocity field ${\bf
b^{nmk}}=(b_r,b_\varphi,b_z)$ corresponding to ${\bf
B^{nmk}}=(B_+,B_-,B_z)$ using the real parts of
respectively  $b_r=B_- + B_+$, $b_\varphi=i(B_- - B_+)$ and $b_z = B_z$ and a fixed sign of
$\lambda$. Some examples of the synthetic velocity fields obtained
for $k/k_0=n=1$ and $m=0,1,2,3$ in the two planes
$\varphi_0=[0,\pi]$ are provided in Fig.\,\ref{fig:velocityfield}.
The recirculation cells, the shear layer and the axi-symmetry of
the experimental time-averaged flow shown
Fig.\,\ref{fig:velocityfield_exp} are well recovered by the
$\varphi$-invariant $m=0$ mode shown in
Fig.\,\ref{fig:velocityfield}(a). We use the boundary condition
explained in Ref.~\onlinecite{boisson_three-dimensional_2011}.
When $m>0$ the velocity fields depend on the azimuthal angle
$\varphi$ and the axi-symmetry is lost. However for each meridian
plane at a fixed [$\varphi_i,\varphi_i+\pi$].
the velocity field shows mirror (anti-)symmetry with respect to
the $r=0$-axis. Even-numbered $m$ velocity fields are found to be
symmetric while odd-numbered are found to be anti-symmetric.
Introducing the space average
$\left\langle\,\right\rangle=\int_{0}^1 rdr\int_{-\pi}^{\pi}
d\varphi\int_{-h}^h dz$ and using the classical properties of the
Bessel functions, the orthogonality of ${\bf \mv^{nmk}}$ writes:
\begin{eqnarray}
\left\langle {\bf \mv^{nmk}}\cdot {\bf \mv^{n'm'k'*}}\right\rangle & 
=&\left\langle { \mv_+^{nmk}} { \mv_+^{n'm'k'*}}+{ \mv_-^{nmk}} { 
\mv_-^{n'm'k'*}} +{ \mv_z^{nmk}} { 
\mv_z^{n'm'k'*}}\right\rangle\nonumber,\\
& =&\left\langle {\bf \mv^{nmk}}\cdot {\bf
\mv^{nmk*}}\right\rangle\delta_{mm'}\delta_{nn'}\delta_{kk'}.
\label{scalarproduct}
\end{eqnarray}
This means that for any field satisfying the decomposition
(\ref{decompoBeltrami}), we have:
\begin{equation}
\left\langle {\bf V}\cdot {\bf
\mv^{nmk*}}\right\rangle=D_{nmk}\left\langle {\bf \mv^{nmk}}\cdot
{\bf \mv^{nmk*}}\right\rangle, \label{projection}
\end{equation}
which provides a simple way to find the projection of the vector
${\bf V}$ on ${\bf \mv^{nmk}}$ by spatial average over the fluid
volume. In the experiment, we however have access to velocity
measurements in one plane only, corresponding to $\varphi=0$ and
$\varphi=\pi$. This precludes the exact instantaneous projection
onto given Beltrami modes. We remedy to these problems
by two methods: i) conditional time averages at a given phase
$m\varphi$. This is explained in Section\,\ref{subsec:vizu}. ii)
Direct comparison with ``synthetic measurements". This is
described in Section\,\ref{subsec:tempbehav}.

These projections, limited to low order Beltrami modes, allow
to reconstruct synthetic velocity field time series in the full 3D
space, accounting for the full spatiotemporal
evolution of the ``organized'' motions of the turbulent flow.


\subsubsection{Spatio temporal spectrum}
In traditional instability analysis, it is customary to draw
space-time diagrams to detect wavelike pattern, see
Refs.~\onlinecite{chate_transition_1987,rubio_spatiotemporal_1989}.
In a turbulent system with a wide range of scales, a natural
generalization is to resort to spatio-temporal spectra. This
approach is routinely used in the field of wave turbulence (see
for a recent review Ref.~\onlinecite{newell_wave_2011}) and led to
computation of dispersion relations, see
Ref.~\onlinecite{cobelli_space-time_2009}. To obtain these spectra from
our experimental PIV measurements, we proceed as follows. From
time series of the azimuthal velocity $v_\varphi(x,z,t)$ in the
meridian plane we compute the time-averaged velocity field ${\bar
v_\varphi}$ and its fluctuations $v'_\varphi(x,z,t)$ defined by
$v'_\varphi=v_\varphi-{\bar v_\varphi}$. From the corresponding
time series, the full space-time (2D in space and 1D in time)
power spectrum $E(k_x,k_z,f)$ is computed, where $k = K/2\pi $
with $K$ the wavevector and $f=\omega/2\pi$ is the frequency. This is done in two steps. First, the
instantaneous spatial (2D) Fourier Transform
$\tilde{v'}_{2D}(k_x,k_z,t)$ for each time $t$ of the time series
is computed. Then, the temporal Fourier Transform of
$\tilde{v'}_{2D}$ using a window size composed of $n$ time steps
is computed. Finally $\tilde{v'}_{3D}(k_x,k_z,f)$ is obtained that
leads to the spatio-temporal power spectrum in 3 dimensions
$E(k_x,k_z,f)$:
\begin{equation}
E(k_x,k_z,f)
=|\tilde{v'}_{3D}|^2
=\left | \int dx\, dz\, dt ~ v_\varphi'(x,z,t) \, e^{i(\omega t + k_x x 
+ k_z z)} \right |^2.
\label{TF}
\end{equation}
Spatial (horizontal and vertical) and temporal resolutions of the
spectra are comprised in the ranges ($1/(2R)$,1/$(2\delta_R)$),
($1/H$,1/$(2\delta_Z)$) and ($f_s/n$,$f_s/2$) respectively.


\subsubsection{Synthetic time series }
Direct comparison of the experimental spectra with synthetic ones,
based on a synthetic velocity fields time series, is then
performed. The latter were obtained with single Beltrami modes of
a given (cylindrical) geometry. For simplicity, we have restricted
our comparison with modes with large-scale poloidal structure
$n=k/k_0=1$ and varying azimuthal structure (variable $m$). As
soon as $m>0$, the axisymmetry is broken. Like in any
rotation-breaking instability, it is natural to expect that the
corresponding perturbation mode will rotate with a frequency $f_r$
(that is a priori small with respect to the impellers rotation
rate $(f_1+f_2)/2$). The corresponding Beltrami modes will then
have the structure:
\begin{equation}
{\bf B}^{1mk_0} \sim \exp(im\varphi + ikz - if_r t).
\label{beltramiperturbation}
\end{equation}
An artificial time series of such a mode is build in the
meridional plane $[x,z]$. This synthetic data is then processed
like the experimental signal to get synthetic space-time power
spectra. Since the perturbation Beltrami mode is periodic (with
$m$ the number of period) in its azimuthal direction, a peak of
energy is located at a frequency $f=mf_r$ as can be seen in
Fig.\,\ref{fig:azimuthalModes}(h). This peak is well resolved by
setting the sampling frequency to a much higher value, typically
$50\times mf_r$. In the sequel, we will refer to the spectra
obtained using the experimental (resp. synthetic) data by $E_{E}$
(resp. $E_S$).

\begin{figure}[h]
\centerline{\includegraphics[width=0.35\textwidth]{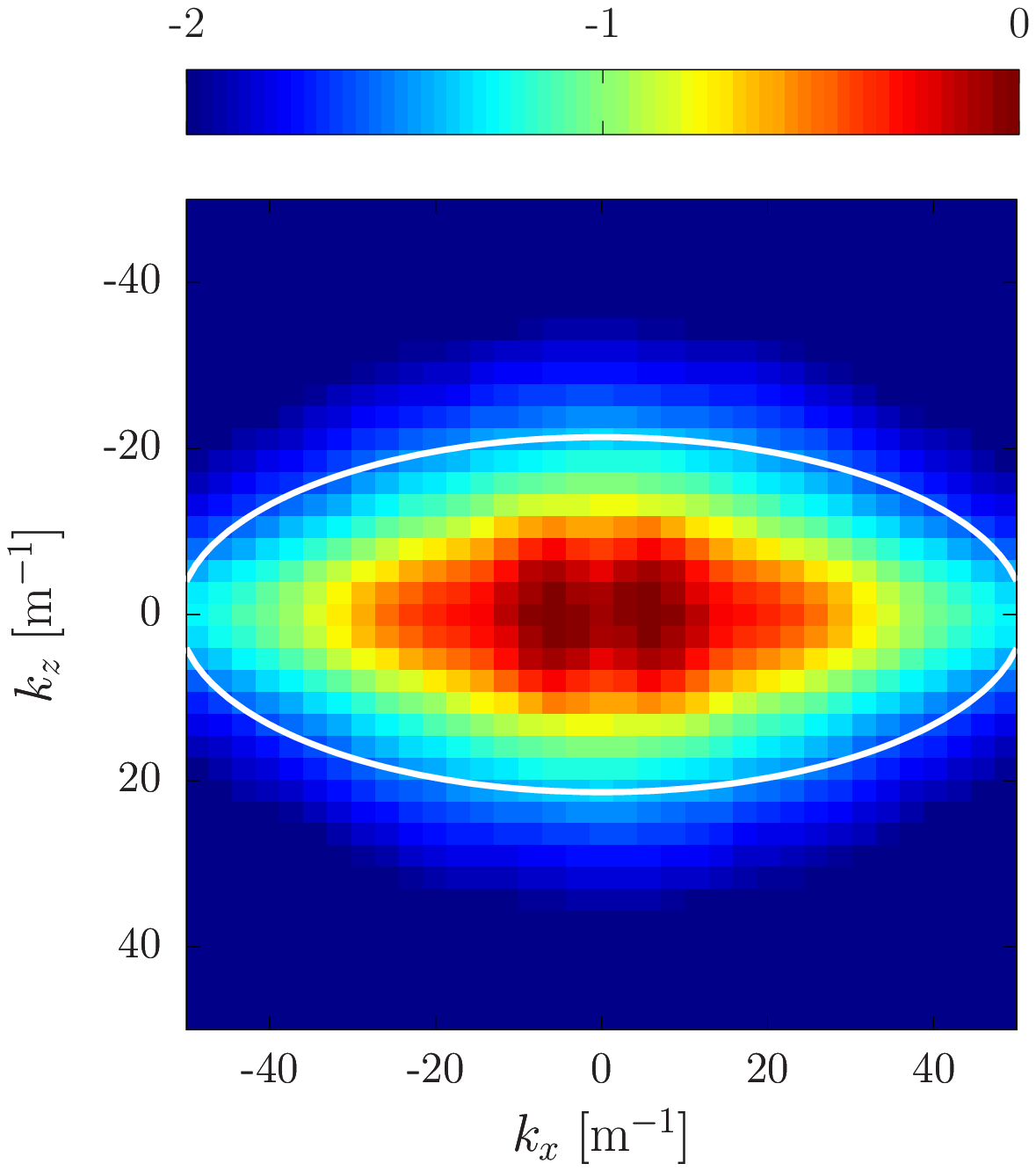}
\hspace{1cm}\includegraphics[width=0.35\textwidth]{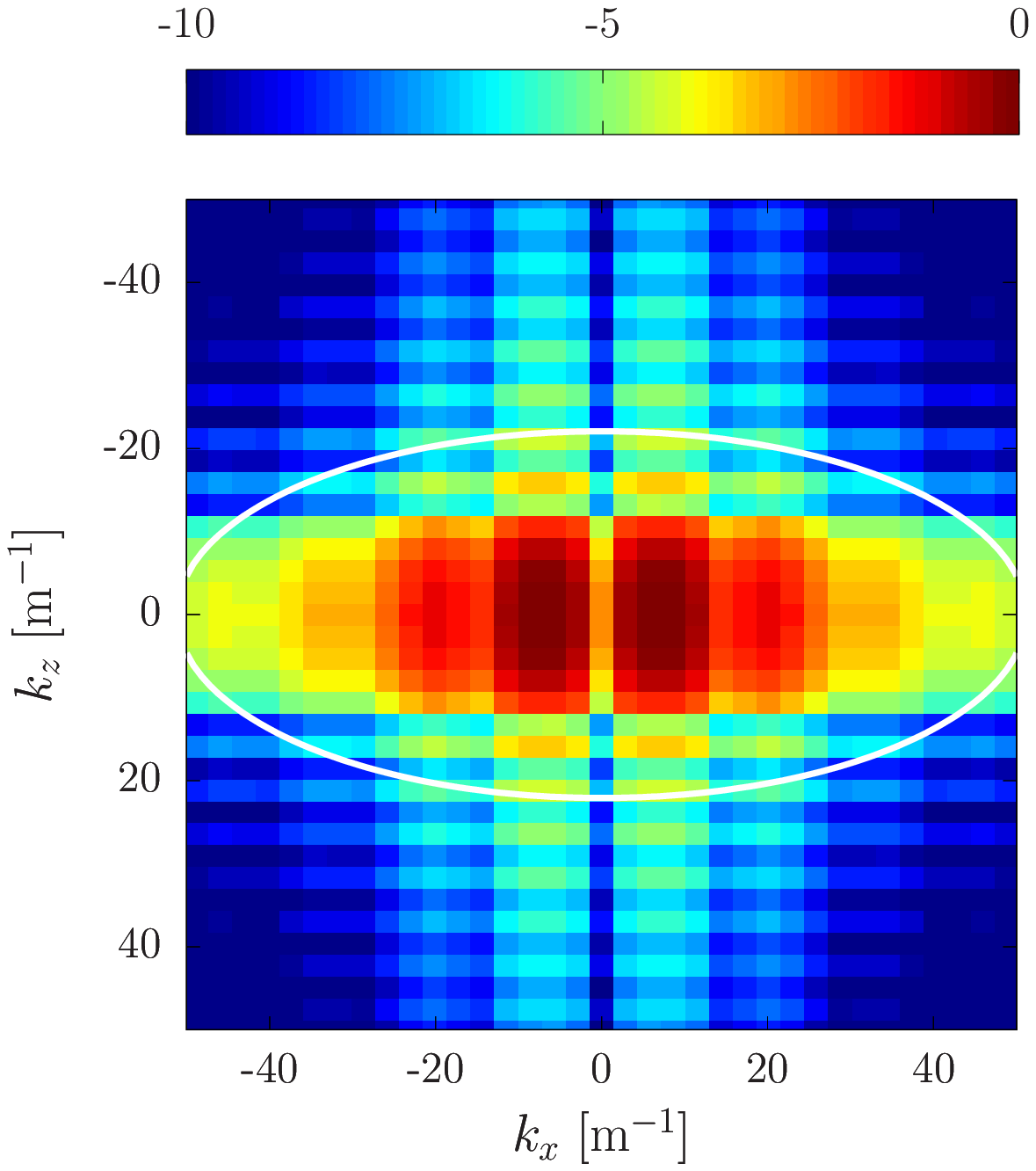}}
     \caption{(Color online) Left, experimental spatial power spectrum
     $E_E(k_x,k_z)$ at $\theta=0$ and $\mathrm{Re}\approx 10^6$, from
     time series from which the flow shown
     Fig.\,\ref{fig:velocityfield_exp}(b) is extracted. White line is
     an ellipsoidal fit of a fixed energy level contour line.
     $w/h=2.5$, with $w$ the width and $h$ the height of the ellipse.
     Right, synthetic spatial power spectrum $E_S(k_x,k_z)$ with
     $n=k/k_0=1$ and $m=3$, corresponding to the flow shown
     Fig.\,\ref{fig:velocityfield}(d) rotating at $f_r=0.2$\,Hz and for
     which $w/h=2.3$. Colors are log scaled.}
     \label{fig:isotropy}
\end{figure}

\begin{figure}[h]
\centering
\begin{tabular}{c}
     \includegraphics[width=0.6\textwidth]{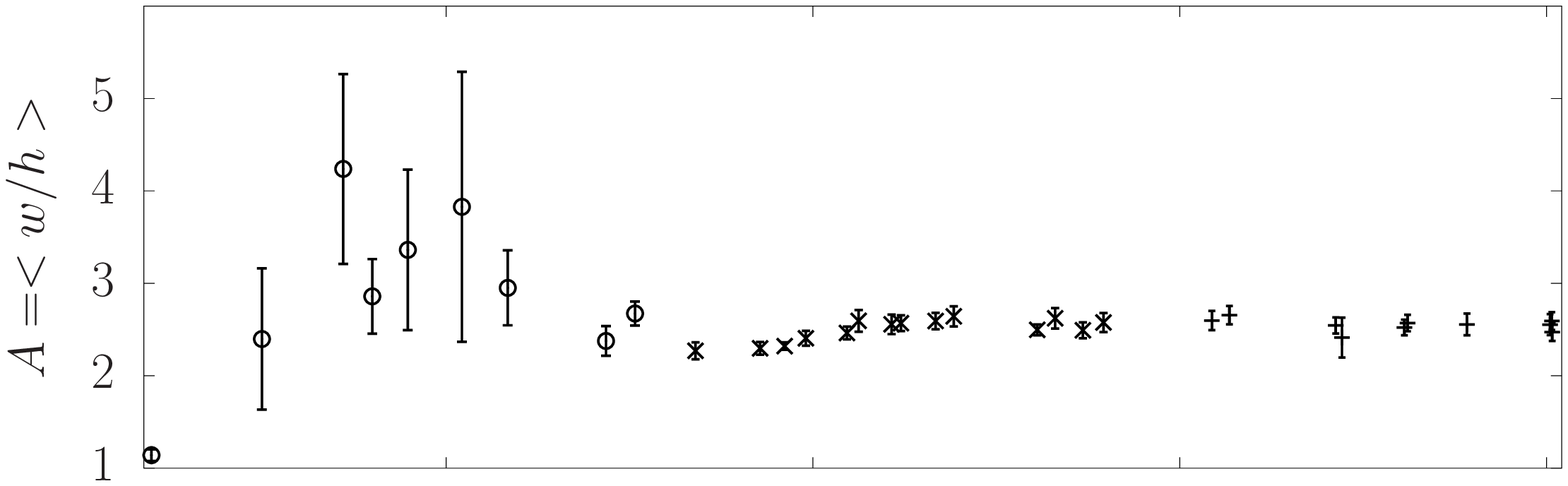} \\
     \includegraphics[width=0.6\textwidth]{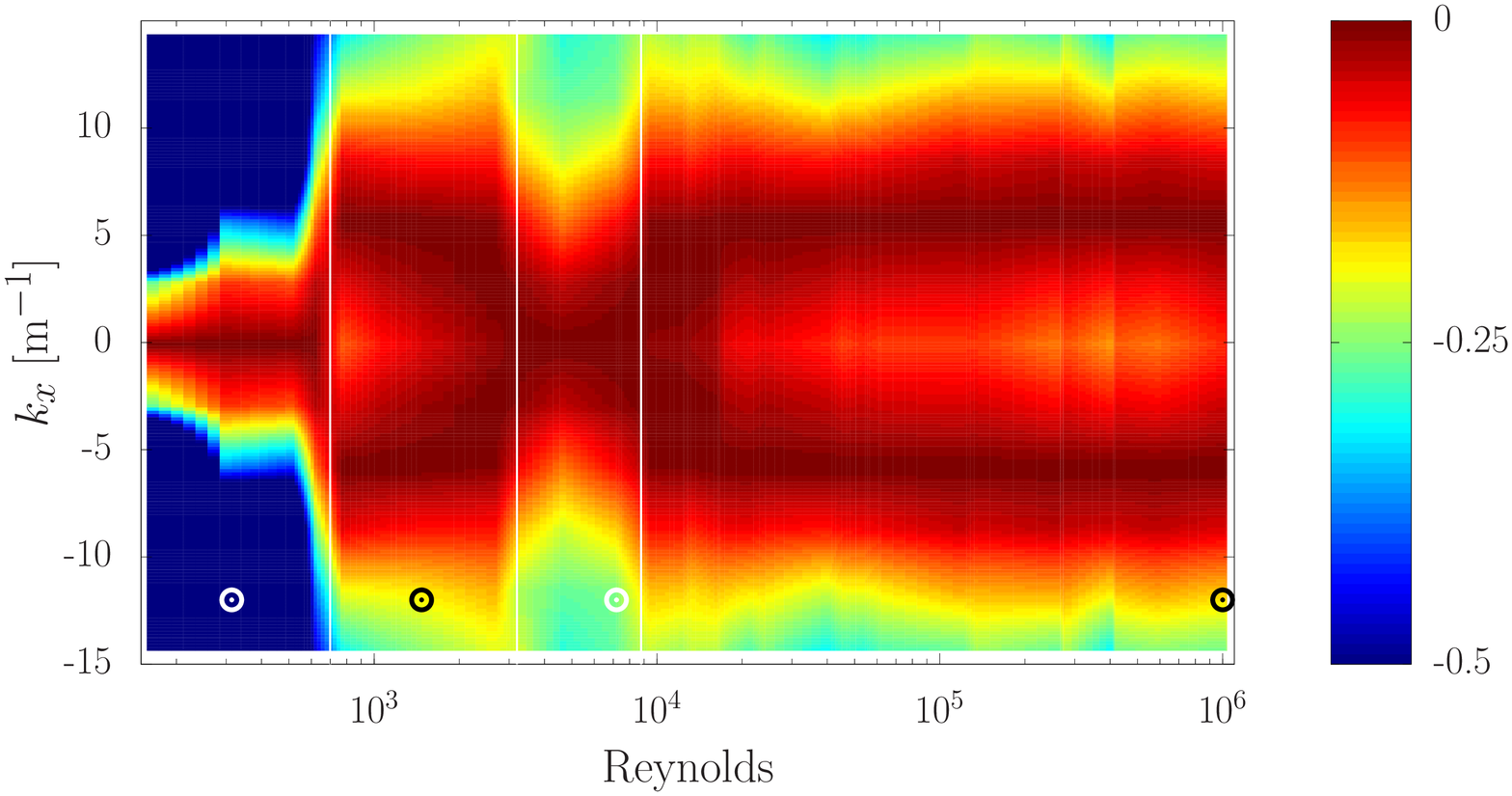} \\
\end{tabular}
\caption{(Color online) Top, anisotropy factor $A=<w/h>$
computed from experimental power spectra as a function of the
Reynolds number (see details in the text and illustration in
Fig.\,\ref{fig:isotropy}). Markers shape ($\circ, \times, +$)
represent the different working fluids. Bottom, experimental
$E_E(k_x)$ at $\theta=0$ as a function of the Reynolds number.
Each spectrum is normalized by its maximum. Colors are log scaled.
Vertical white lines mark strong transitions in the $E_E(k_x)$
spectrum which highlight parity changes in the structure of the
experimental flow. Black and white circles, see
Fig.\,\ref{fig:azimuthalModes}.} \label{fig:isotropyRe}
\end{figure}

\section{Results}
\label{sec:results}
\subsection{Isotropy -- Evidence for the Eckhaus instability}
\begin{figure}[htb]
\centering
\begin{tabular}{cccc}
 
\includegraphics[width=0.26\textwidth]{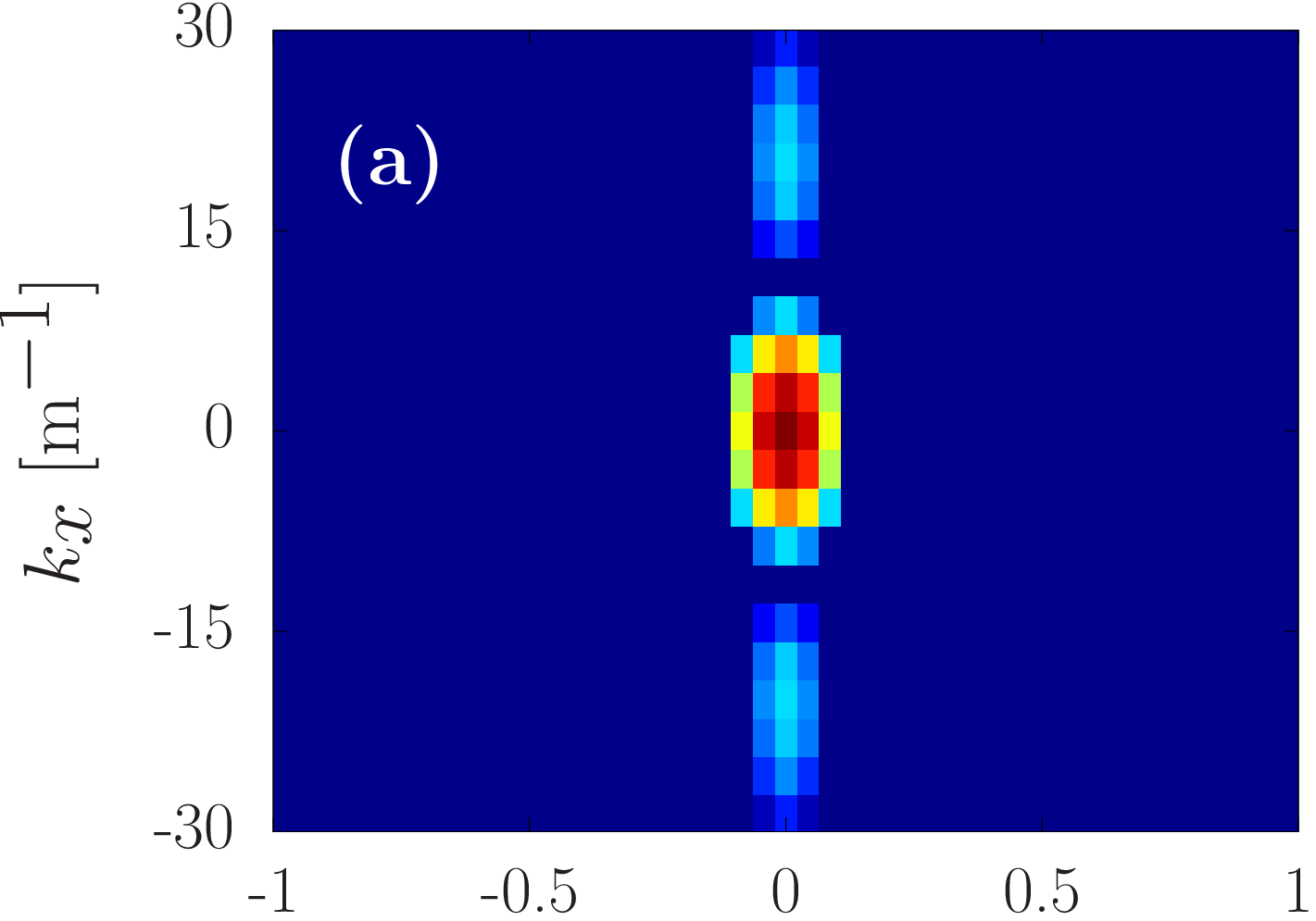} &
\includegraphics[width=0.23\textwidth]{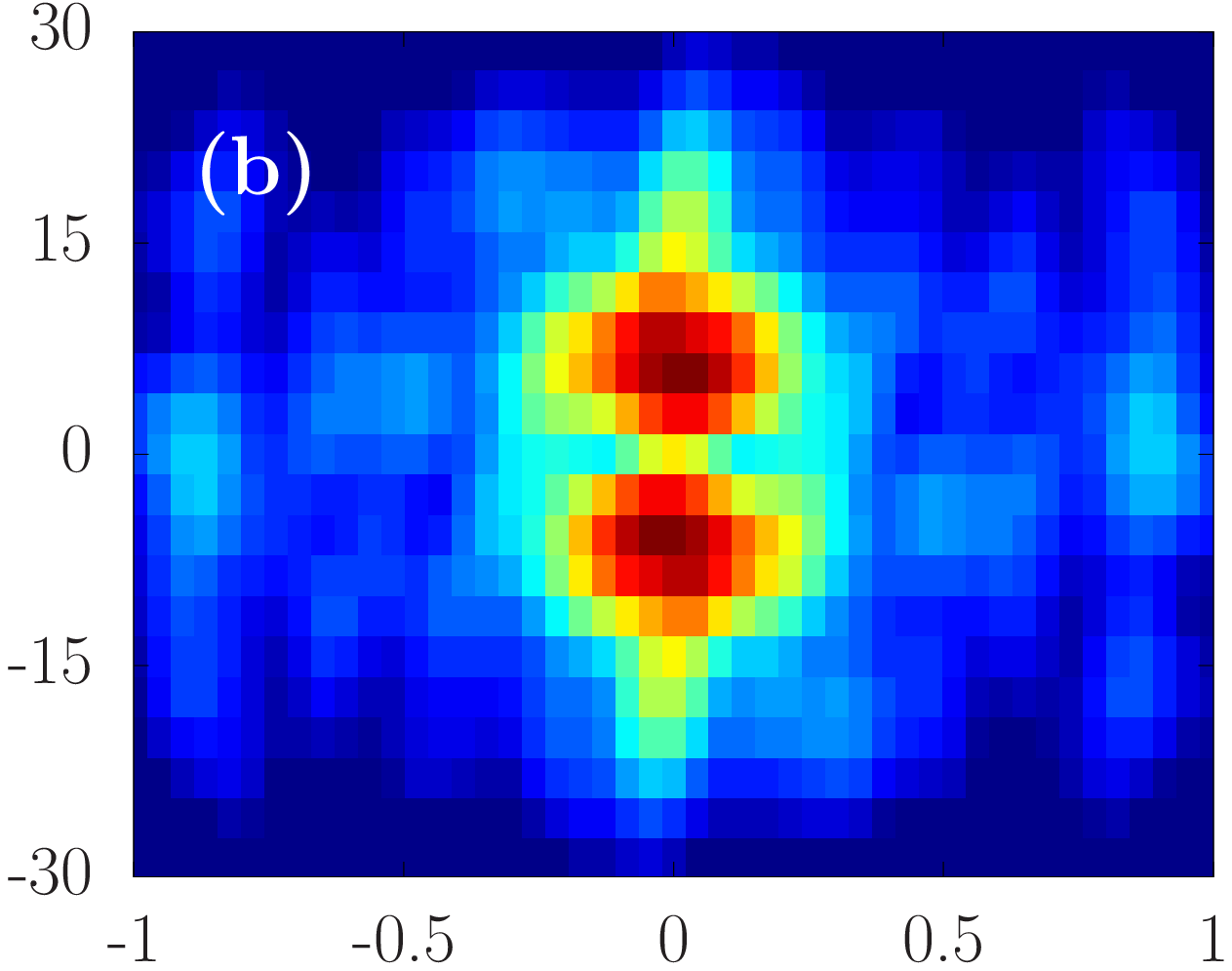} &
\includegraphics[width=0.23\textwidth]{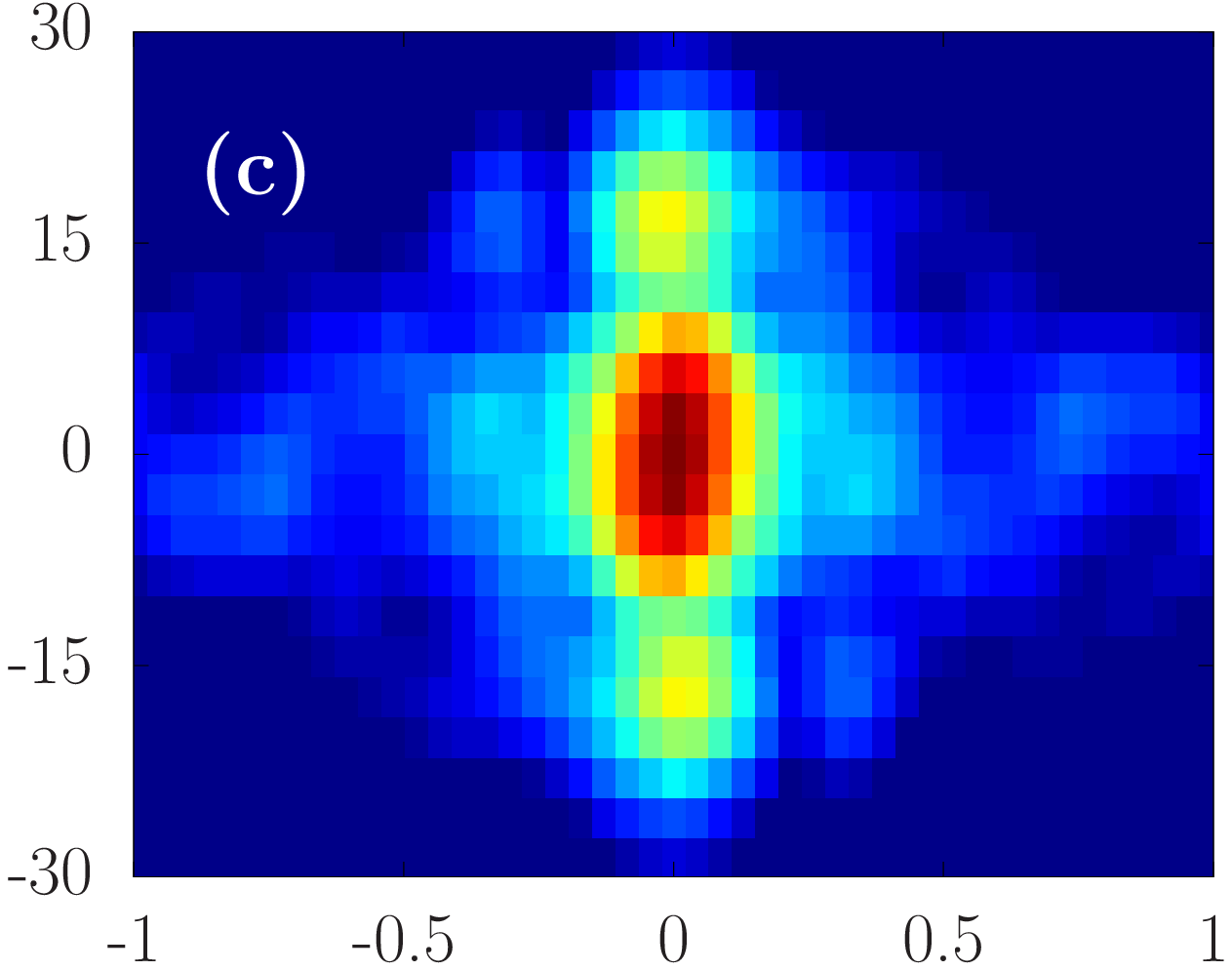} &
   \includegraphics[width=0.23\textwidth]{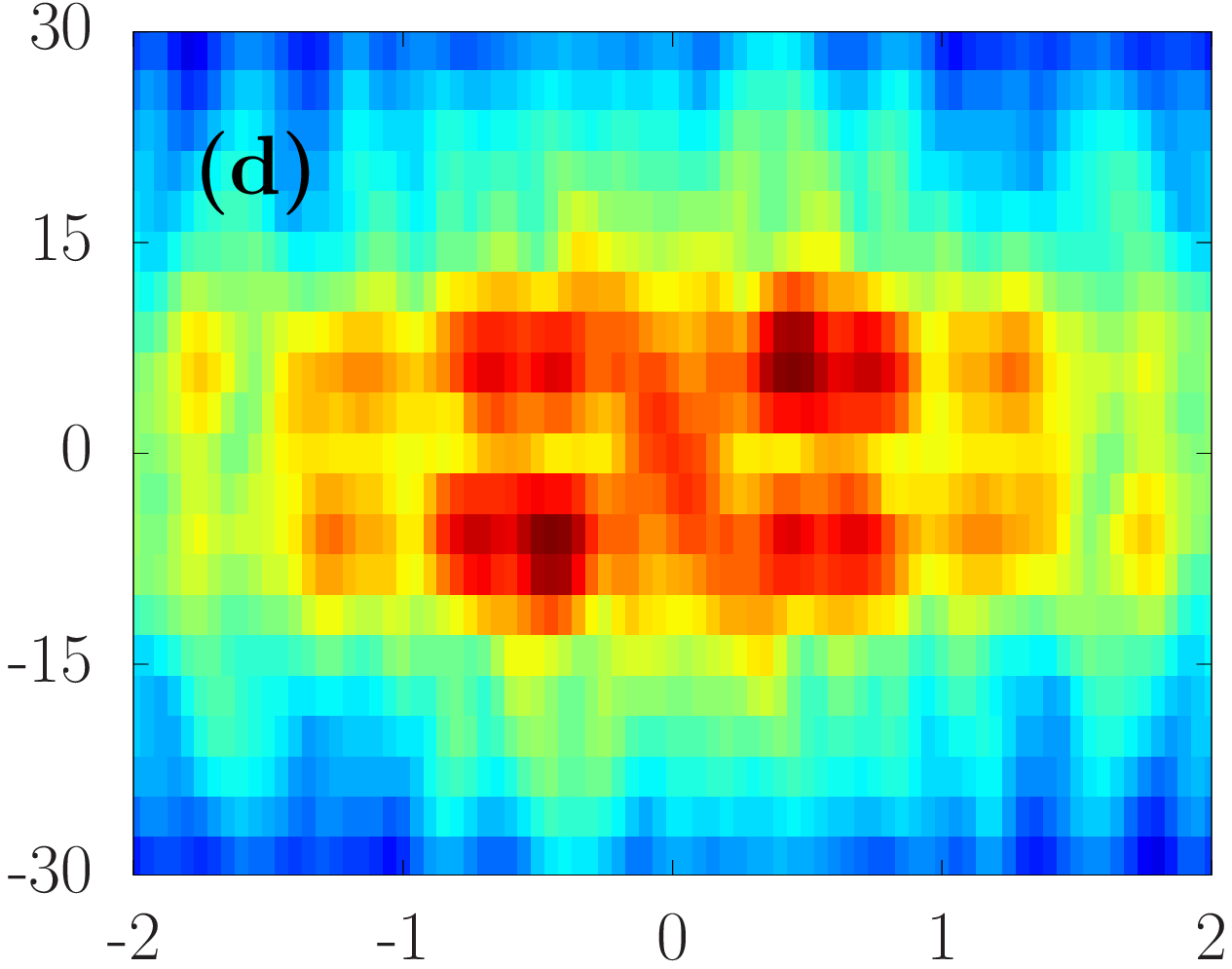} \\
   \includegraphics[width=0.26\textwidth]{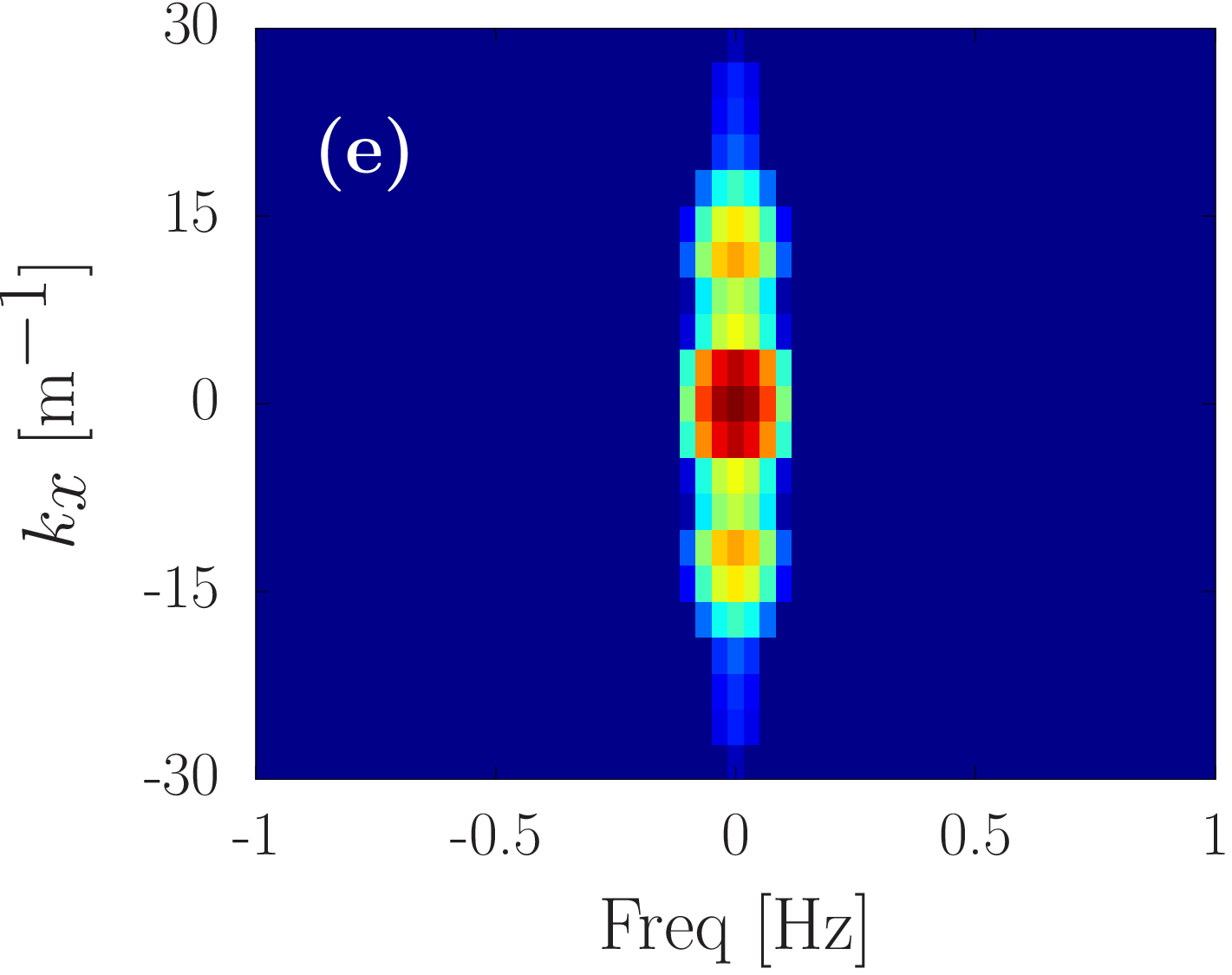} &
   \includegraphics[width=0.23\textwidth]{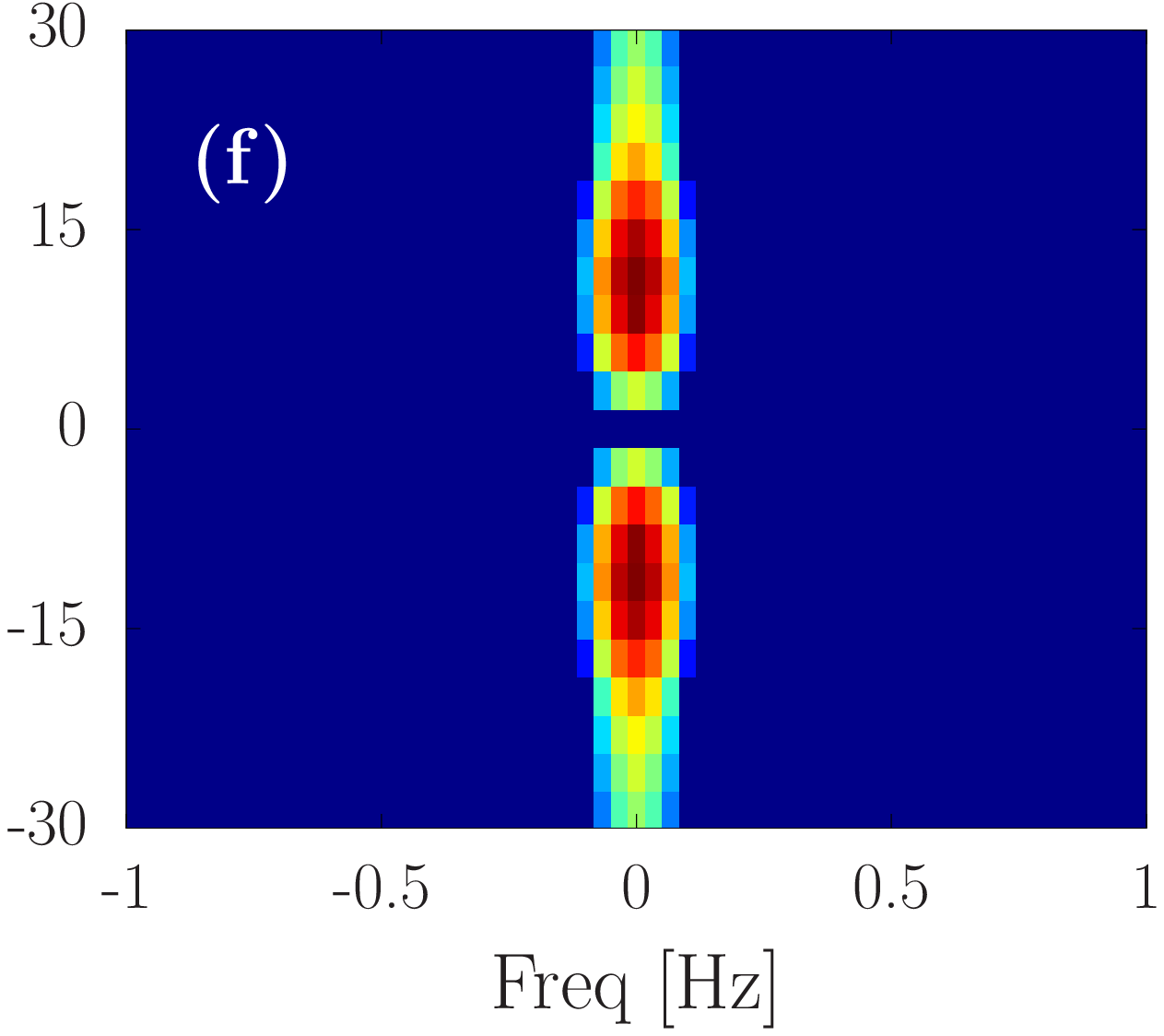}  &
   \includegraphics[width=0.23\textwidth]{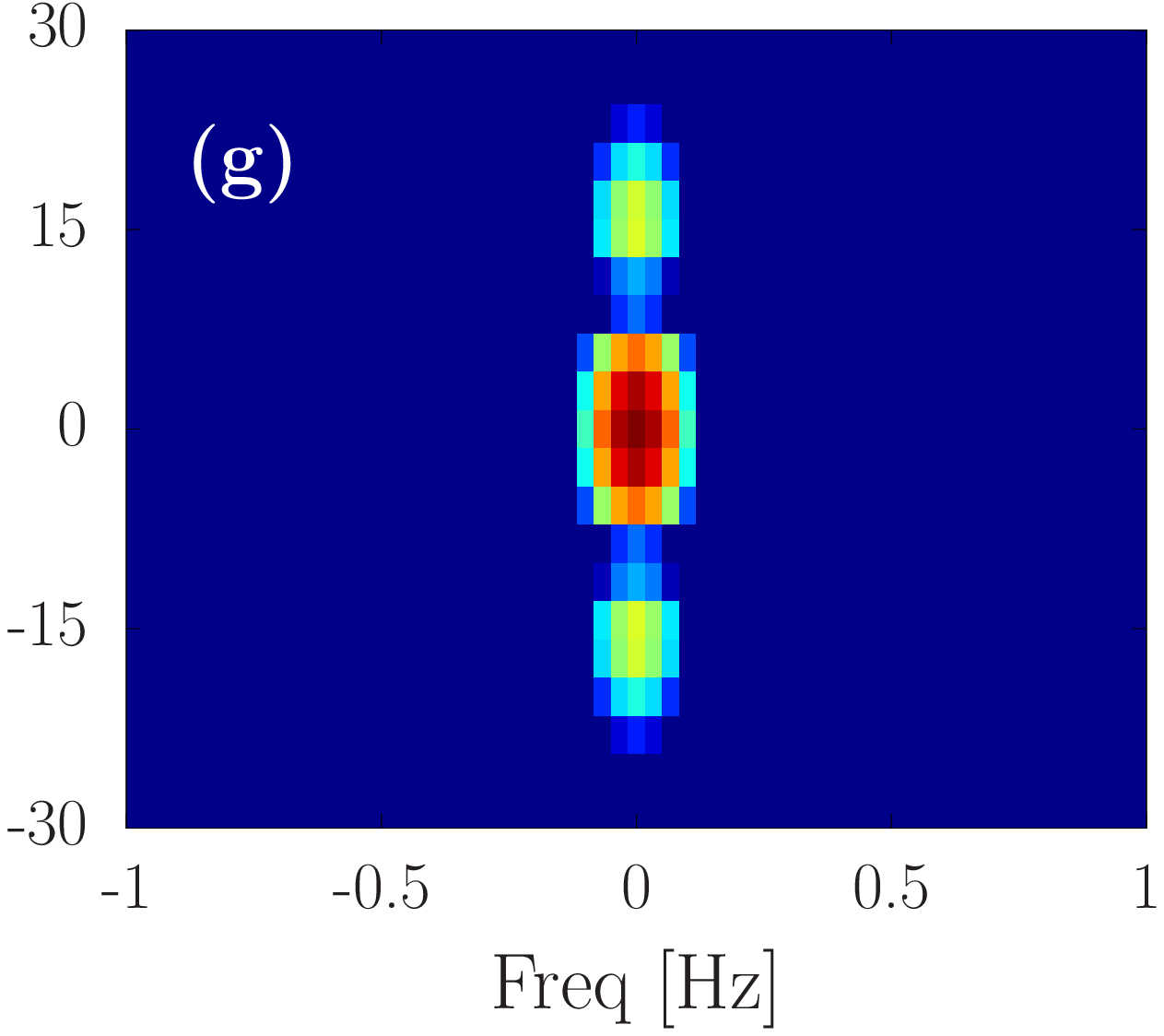} &
   \includegraphics[width=0.23\textwidth]{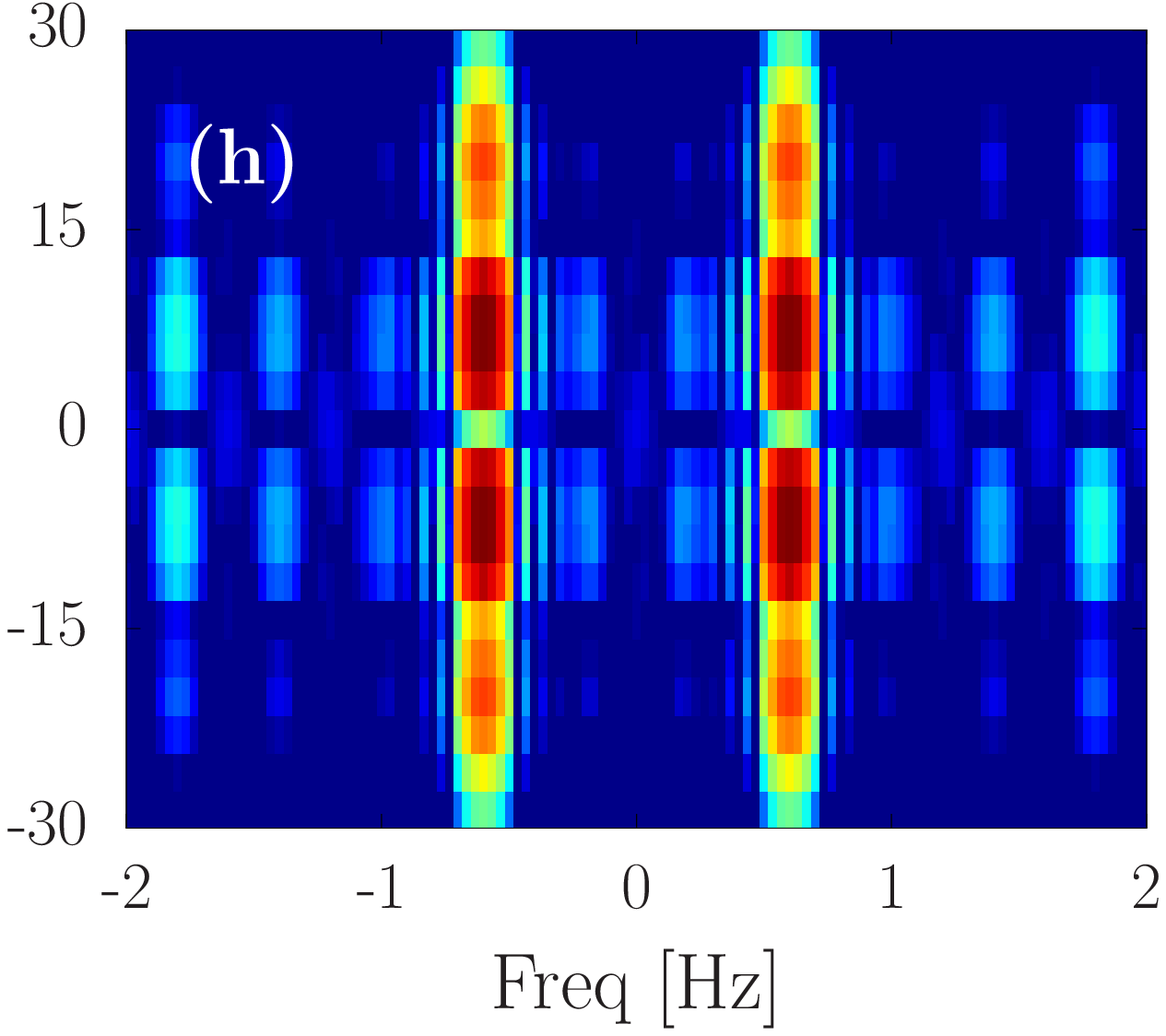}  \\
\end{tabular}
\caption{(Color online) Top, experimental spatio-temporal spectrum
$E_E(k_x,f)$ at $\theta=0$ for $\mathrm{Re}=314$ (a), 1473 (b),
7200 (c) and $10^6$ (d), corresponding to circles in
Fig.\,\ref{fig:isotropyRe}. Colors are log scaled in the range
$[-1.5, 0]$. Bottom, synthetic spatio-temporal spectrum
$E_S(k_x,f)$, using  $n=k/k_0=1$ and $f_r=0.01$\,Hz for $m=0$ (e),
1 (f), 2 (g) and $f_r=0.2$\,Hz for $m=3$ (h). Colors are log
scaled in the range $[-2,0]$ for (e-g) and $[-6,0]$ for (h). Note
the frequency axis range change for (d) and (h).}
\label{fig:azimuthalModes}
\end{figure}

A first characterization of the flow perturbation independently of
its dynamics can be found by considering the spatial spectrum
$E(k_x,k_z)$, obtained through the integration of the 3D spectrum
$E(k_x,k_z,f)$ over all temporal frequencies. An example for an
experimental flow at $\mathrm{Re}\approx 10^6$ is provided in
Fig.\,\ref{fig:isotropy}. The energy is mainly concentrated along
the $x$-axis at $k_z\approx 0$ inside an ellipsoidal region,
pointing out the anisotropic structure of the flow. For useful
interpretation, we have compared this isotropy measurements with
synthetic data. In Fig.\,\ref{fig:isotropy}, we also show an
example obtained with $n=k/k_0=1$, $m=3$ and $f_r=0.2$\,Hz
synthetic velocity fields time-series. For this field, the overall
shape of the spatial spectrum is close to the experimental one.
Both spatial spectrum $E_E(k_x,k_z)$ and $E_S(k_x,k_z)$ in
Fig.\,\ref{fig:isotropy} display 2 distinct maxima. They are found
around $k_z=0$ for two values of $k_x$ symmetric with respect to
$k_x=0$. As we can see in Fig.\,\ref{fig:velocityfield} using the
synthetic velocity fields or in Fig.\,\ref{coherentstructure}
using the projection of the experimental fields onto the Beltrami
modes, the $x$-axis traces the parity of $m$. This parity is
reproduced through the spectral analysis in this way: in the
even-numbered (resp. odd-numbered) case, the symmetry (resp.
anti-symmetry) of the flow with respect to $r=0$ leads to a
fundamental wavelength of roughly the size (resp. half of the
size) of the vessel. The wavenumbers $k_x \simeq \pm 1/(2R)$
corresponding to the maxima of energy in the symmetric case
approach the resolution limit of the Fourier transform and
practically appear as $k_x \simeq 0$ (see
Figs.\,\ref{fig:azimuthalModes}(e)(g)). On the contrary, the
anti-symmetric case leads to two distinguishable symmetric peaks
at $k_x\approx \pm 1/R$ (see
Figs.\,\ref{fig:azimuthalModes}(f)(h)). Experimentally these
different patterns are reproduced using the Reynolds number as a
driving parameter. Figs.\,\ref{fig:azimuthalModes}(a)(c) at
$\mathrm{Re}=714 $ and 7200 show the symmetric (even) case whereas
Figs.\,\ref{fig:azimuthalModes}(b)(d) at $\mathrm{Re}=1473 $ and
$10^6$ show the anti-symmetric (odd) case. Consequently the flow
is experiencing successive transitions in parity. To study more
precisely the evolution of this pattern with the Reynolds number,
one can concentrate on the 1D-spatial spectrum $E_E(k_x)$,
obtained through integration of $E_E(k_x,k_z)$ over all $k_z$.
This was done for all values of $\mathrm{Re}$, resulting in
Fig.\,\ref{fig:isotropyRe}. One observes clear transitions at
$\mathrm{Re}=700$, $3200$ and $8800$, where the maximum of the
1D-spatial spectrum shifts back and forth between $k_x=0$ and two
symmetric values $k_x=\pm 7$\,m$^{-1}$. Therefore, the kind of
transitions observed in Fig.\,\ref{fig:isotropyRe} is suggestive
of successive azimuthal changes in parity (even-odd-even-odd).
This means that the flow is composed of a large scale coherent
structure which is experiencing an Eckhaus type instability in its
azimuthal direction starting from the axi-symmetric time-averaged
flow which is a $m=0$ mode.

To bring some clues  to the upper modes where $m>0$  we resort to
a different analysis of the spatial spectrum. 
The contour line at a fixed iso-energy level was fitted using a centered ellipsoidal curve,
$(x/w)^2+(z/h)^2=1$, with the width $w$ and the height $h$ as
fitting parameters. 
 As a result $w/h$ is a measure of the anisotropy of the
flow. 
In order to obtain error bars, the ratio $w/h$ was computed for various iso-energy levels corresponding to different fractions of the total energy of the spectra, from 65 to 90\%
(resp. 75\% to 99\%) for $E_E$ (resp. for $E_S$). The mean value
$A=<w/h>$ and the standard deviation are finally computed to
evaluate both the anisotropy and the error. Using the synthetic data, varying $m$ leads to
different values of $A$ which is moreover found not to depend on
$f_r$.  Specifically, $A=2.3\pm 0.3$, for
$m=0$. For $m=1$, $A=5.2\pm 2.4$ reach its maximum. Increasing
$m$, $A$ tends to decrease, yielding for $m=2$, $3$ and $4$ to
$A=2.9\pm 0.8$, $2.4\pm 0.3$ and $2.3\pm 0.3$ respectively.
We have then studied the experimental evolution of the anisotropy ratio with the Reynolds number. This
is shown in Fig.\,\ref{fig:isotropyRe}. One sees the error bar can be large in particular below
$Re=5000$. Consequently this approach is not precise enough to
determine at which Reynolds number the transitions occurred.
However we clearly distinguish different regimes. The anisotropy
starts from a value of about $1.1$ at very low Reynolds number
where the amplitude of the fluctuations reach the resolution limit
of our PIV. Increasing the Reynolds number, $A$ increases up to
$2.7\pm 0.7$ corresponding to the laminar axi-symmetric ($m=0$)
case where $A=2.3$ or to the time-averaged flow at any Reynolds
number. The amplitude of the anisotropy then reaches a maximum
($4.2\pm 1$) at $\mathrm{Re}=800$. This last
value is compatible with the odd-numbered $m=1$ mode. At
approximately $\mathrm{Re}=5000$, $A$ reaches a minimum at $2.3\pm
0.1$. Which seems not compatible with even-numbered $m=2$ mode but
could be a $m=4$ mode. Finally for Reynolds number larger than
$10^4$, $A$ shows a plateau around $A=2.5$, which value is
compatible with the odd-numbered $m=3$ mode.

\subsection{Temporal behaviour}
\label{subsec:tempbehav}

A symmetry breaking generally coincides with a propagative
instability.\cite{coullet_instabilities_1990} To check this
hypothesis and try to measure the rotating frequency $f_v$ of the
azimuthal structure of the turbulent flow, we used the
2D-spatio-temporal spectrum $E_E(k_x,f)$ which is computed through
the integration over $k_z$ of the 3D-spectrum $E(k_x,k_z,f)$.
$E_E(k_x,f)$ is then compared to the spectrum $E_S(k_x,f)$
obtained using a synthetic Beltrami flow for different values of
$m$. This is done in Fig.\,\ref{fig:azimuthalModes} using
different Reynolds numbers. The rotation frequency $f_r$ of the
synthetic flows is adjusted to get the best resemblance with the
experimental field which is guessed to rotate at a $f_v$
frequency.
At $\mathrm{Re}=314$, the experimental spectrum is found very
close to $E_S(k_x,f)$ with $m=0$. This result can be extended to the
time-averaged flow spectra (not shown) at any Reynolds number
since it is similar to the laminar instantaneous flow. Increasing
the Reynolds number to $\mathrm{Re}=1473$ (resp.
$\mathrm{Re}=7200$) in a region where one (resp. two) Eckhaus
bifurcation(s) has occurred, we observe a strong similarity with
corresponding $E_S(k_x,f)$ with $m=1$ (resp. $m=2$) with a small
but non-zero rotation frequency $f_r=0.01 $\,Hz. For the last two
cases the rotation frequency $f_v$ of the azimuthal structure of
the flow shall be non-zero. However, it reveals to be smaller than
our spectral resolution, typically $0.01$\,Hz,  and therefore
remains unmeasurable.
Increasing the Reynolds number the energy is found to remain
mainly concentrated at a 0 frequency up to $\mathrm{Re}=4\times
10^4$. In the vicinity of this last $\mathrm{Re}$ the energy
becomes distributed on a wider range of frequencies, up to
1.5\,Hz. However no particular frequency is selected. Increasing
further the Reynolds number, peaks at non-zero frequencies come
up, as shown for $\mathrm{Re}=10^6$. These peaks are the signature
of a fixed rotation frequency at $f_v=0.7$\,Hz.  In this case, the
experimental spectrum is regained by $E_S(k_x,f)$ with $m=3$ and
$f_r=0.2$\,Hz. Overall, our results are compatible with a rotating
frequency bifurcating from a small but non-zero value around
$\mathrm{Re}=4\times 10^4$.


\subsection{Rotation direction}
\begin{figure}[h]
\centering
\begin{tabular}{cc}
   \includegraphics[width=0.26\textwidth]{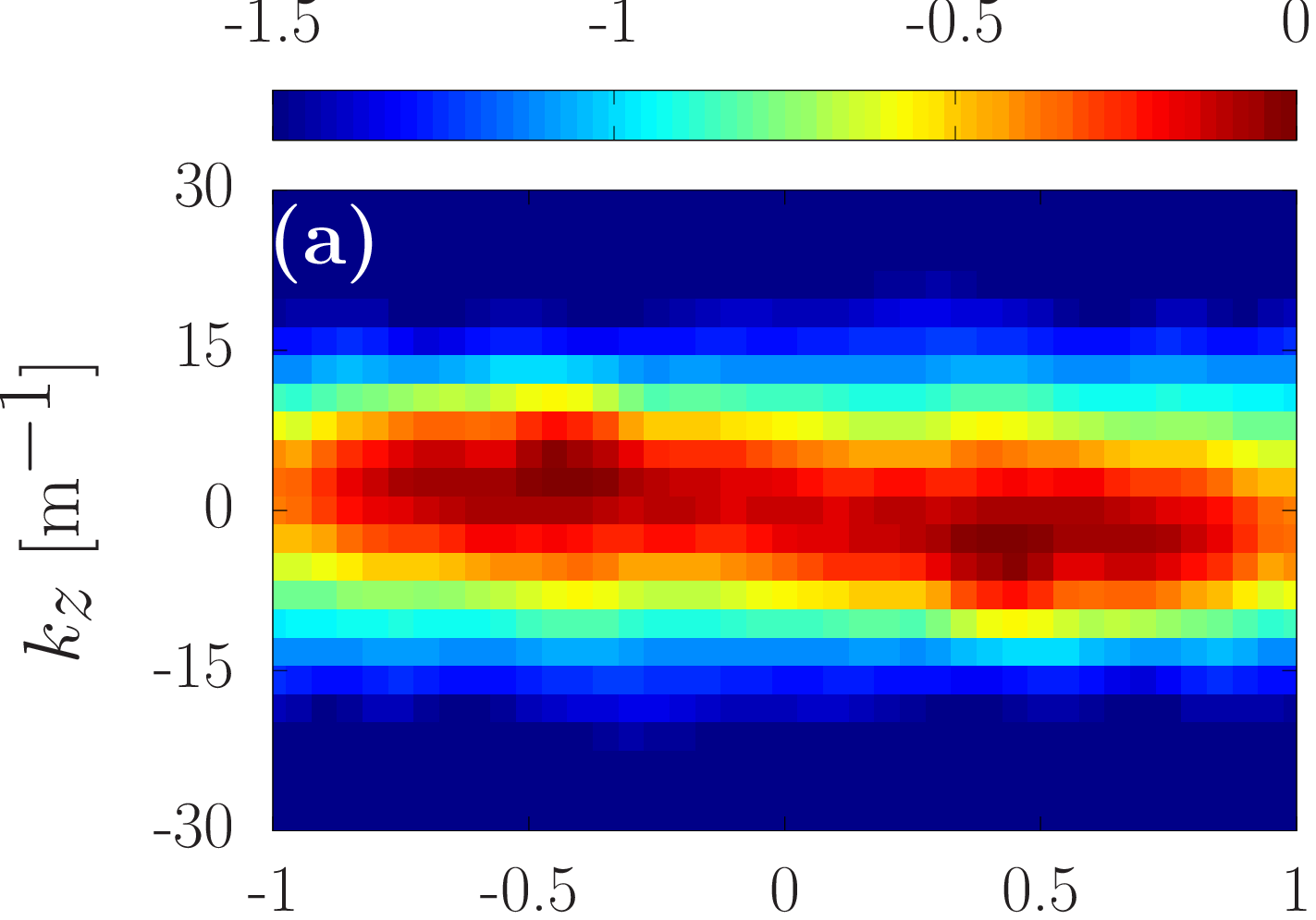} &
   \includegraphics[width=0.23\textwidth]{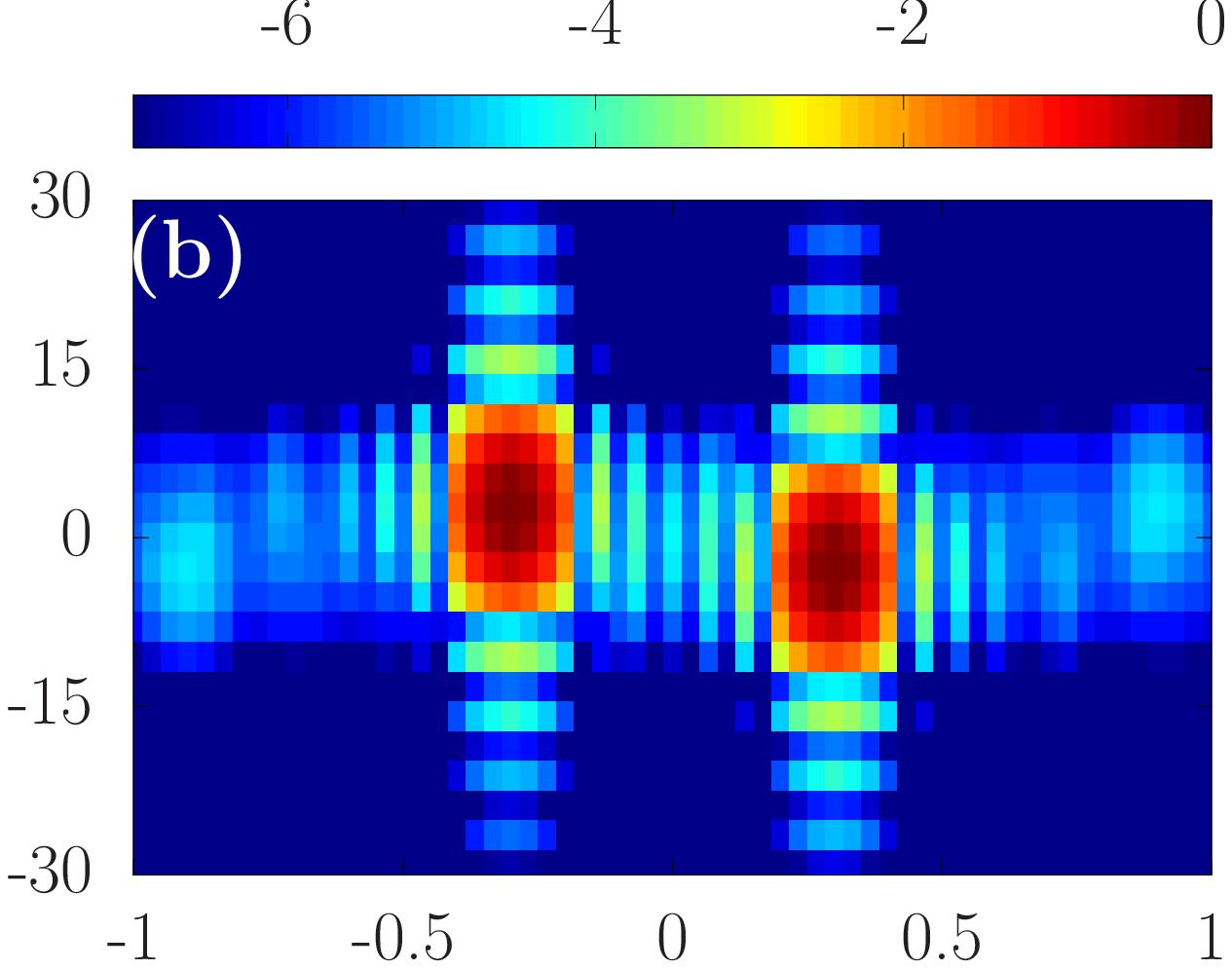} \\
\includegraphics[width=0.26\textwidth]{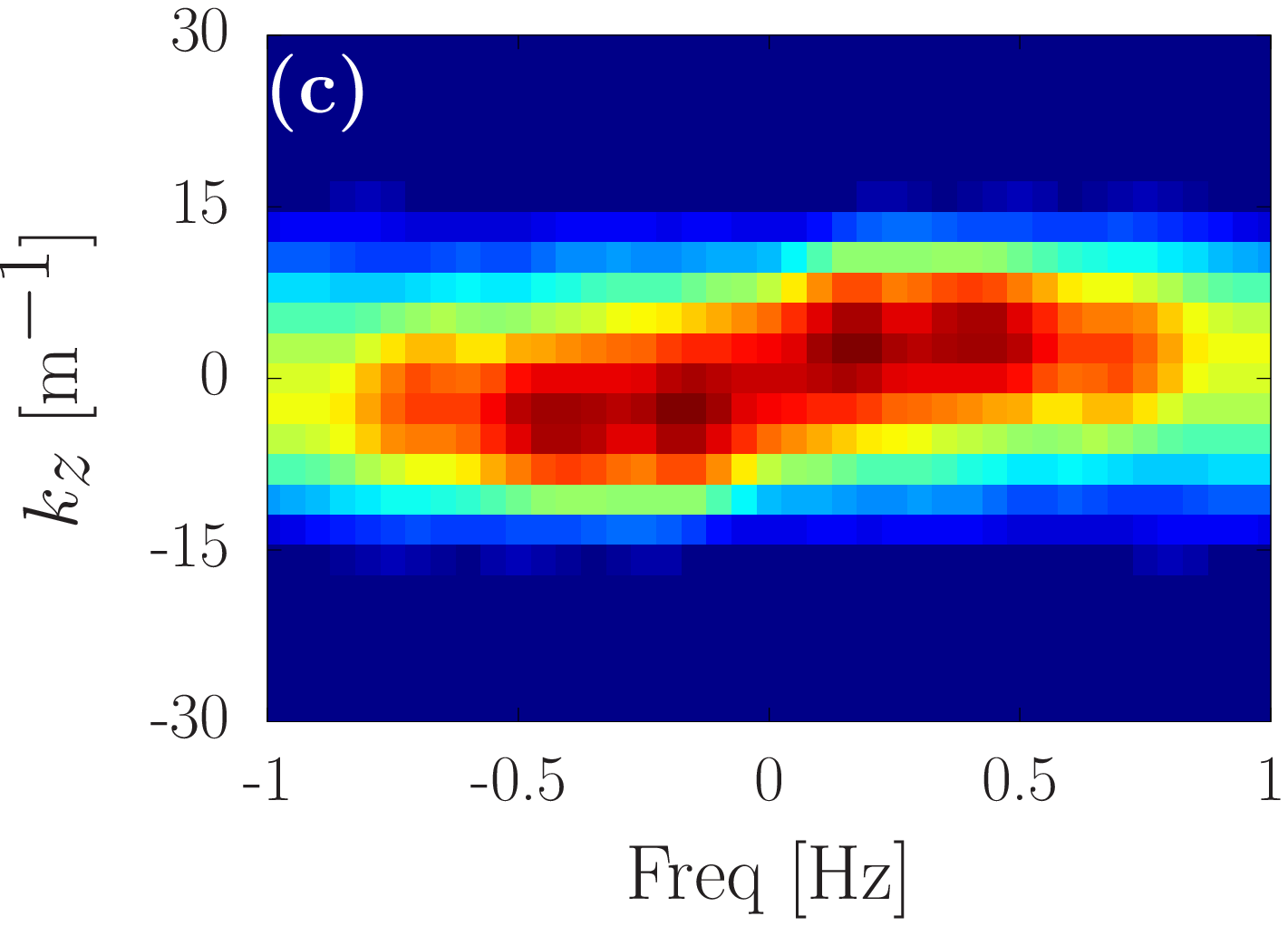} &
   \includegraphics[width=0.23\textwidth]{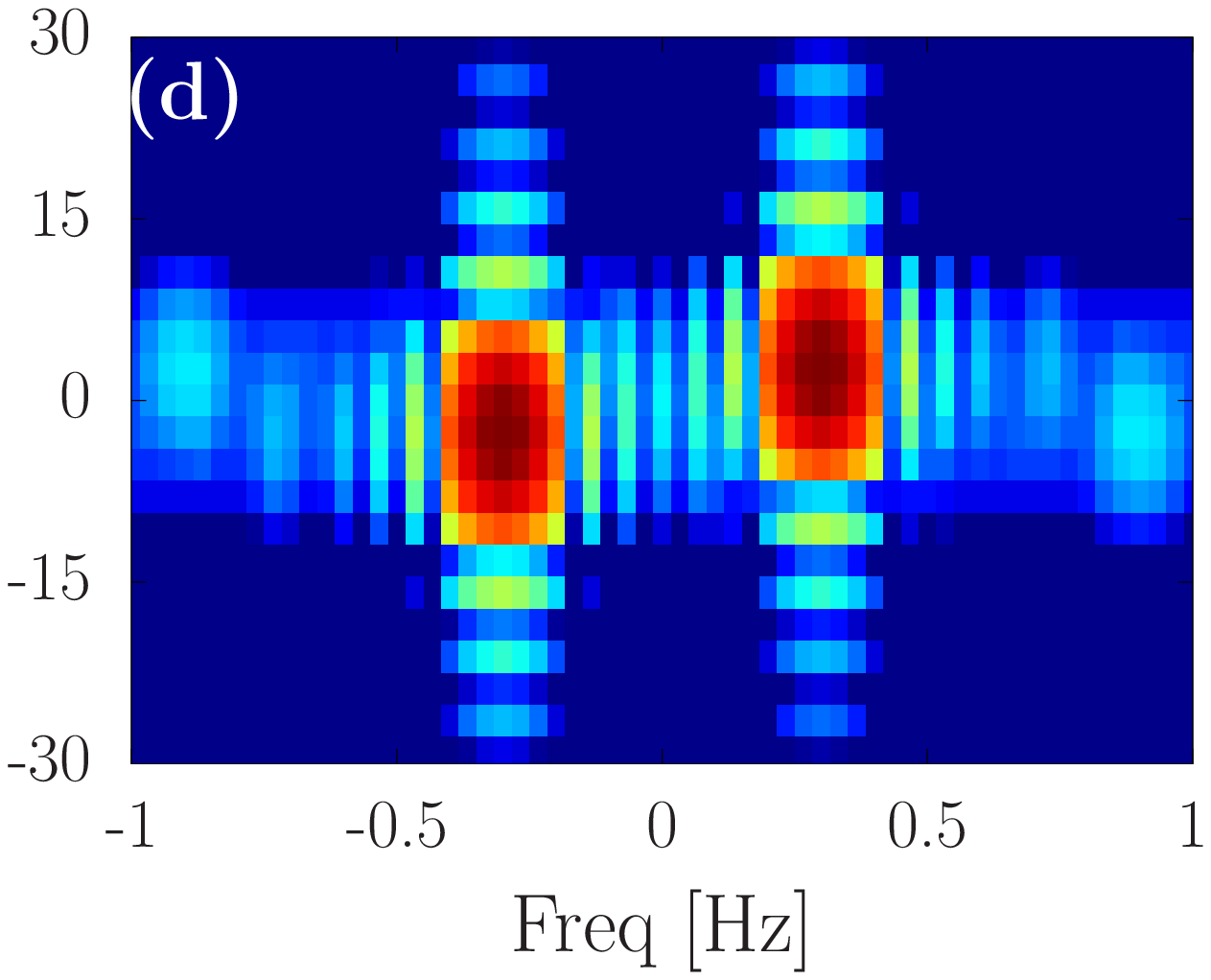} \\
\end{tabular}
\caption{(Color online) Left, experimental spatio-temporal
spectrum $E_E(k_z,f)$ for (a) $\theta=0$, $\mathrm{Re}\approx
10^6$ and (c) $\theta = -8\times 10^{-3}$, $\mathrm{Re}\approx
8\times 10^5$. Right, synthetic spatio temporal spectrum
$E_S(k_z,f)$ with $n=k/k_0=1$, $m=3$ and using (b) $f_v=0.1$\,Hz
and (d) $f_v=-0.1$\,Hz. Colors are log scaled.}
\label{fig:rotationsens}
\end{figure}

At large enough Reynolds number, we have seen that a $m=3$
perturbation of the ``basic state" of the flow is established with
a non-zero rotating frequency. In the perfectly symmetric case
there is however no general argument imposing the direction of
rotation of the pattern. Experimentally this should be the case at
$\theta=0$. However our setup was not perfectly symmetric what led
to observe the same direction of rotation for all experiments. A
slight shift of $\theta$ was then introduced to compensate this
lack of symmetry. Using this adjustment we observed for each
experiment one or the other direction of rotation with an equal
probability.  The direction of rotation is best seen by focusing
on the vertical direction, i.e. looking at the 2D-spatio-temporal
spectrum $E_E(k_z,f)$, like shown in Fig.\,\ref{fig:rotationsens}.
One observes either an upward ($k_z>0$ when $f>0$) or a downward
($k_z<0$ when $f>0$) tilt of the spatio-temporal spectrum. This
can be modelled by considering a synthetic Beltrami spectrum
corresponding to $m=3$ with two opposite rotation directions,
leading to an upward or downward tilt of the spectrum (also shown
Fig.\,\ref{fig:rotationsens}).

\subsection{Link with the shear layer dynamics}
\begin{figure}[h]
\centering
\begin{tabular}{cc}
   \includegraphics[width=0.347\textwidth]{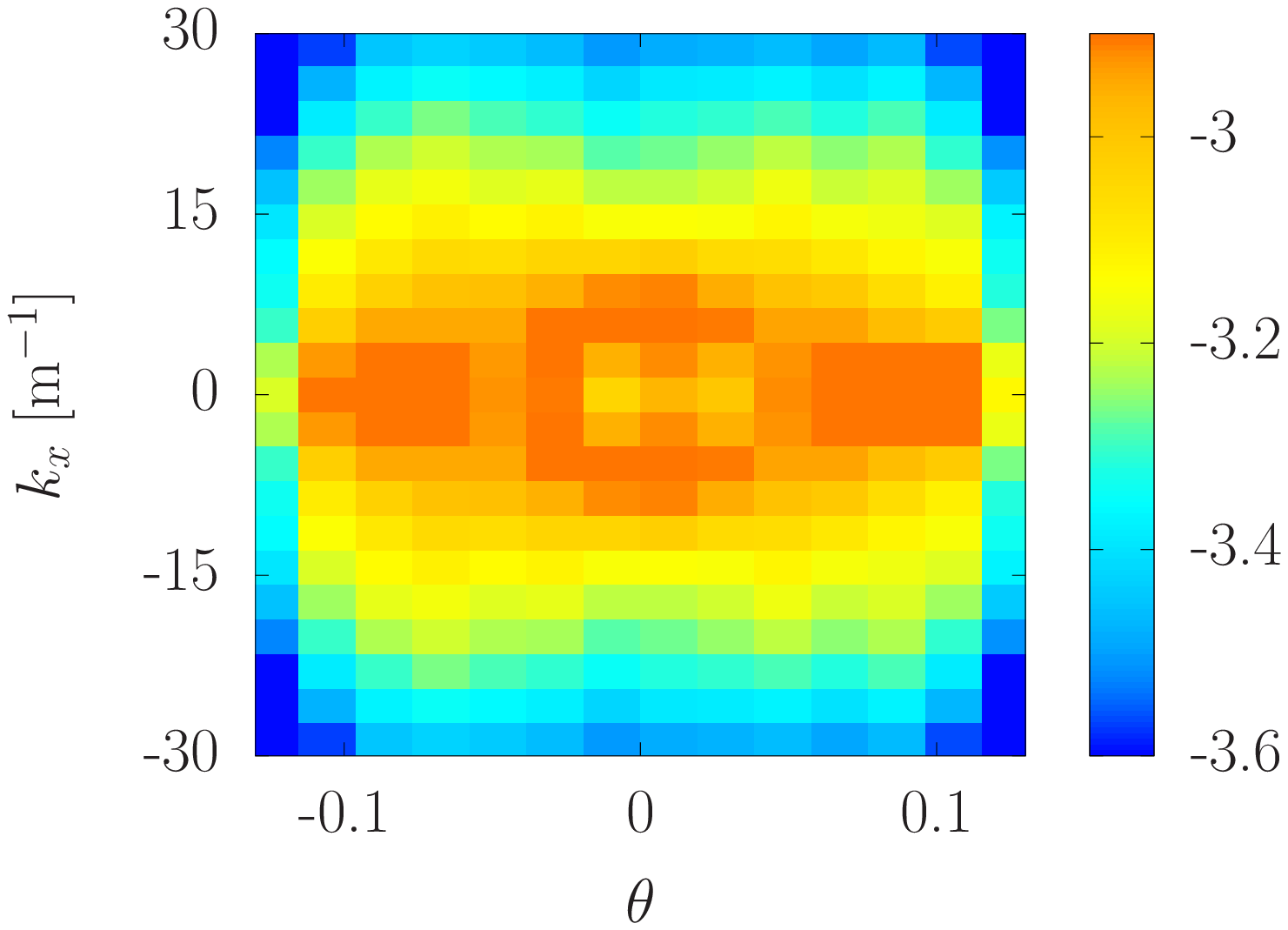} &
   \includegraphics[width=0.31\textwidth]{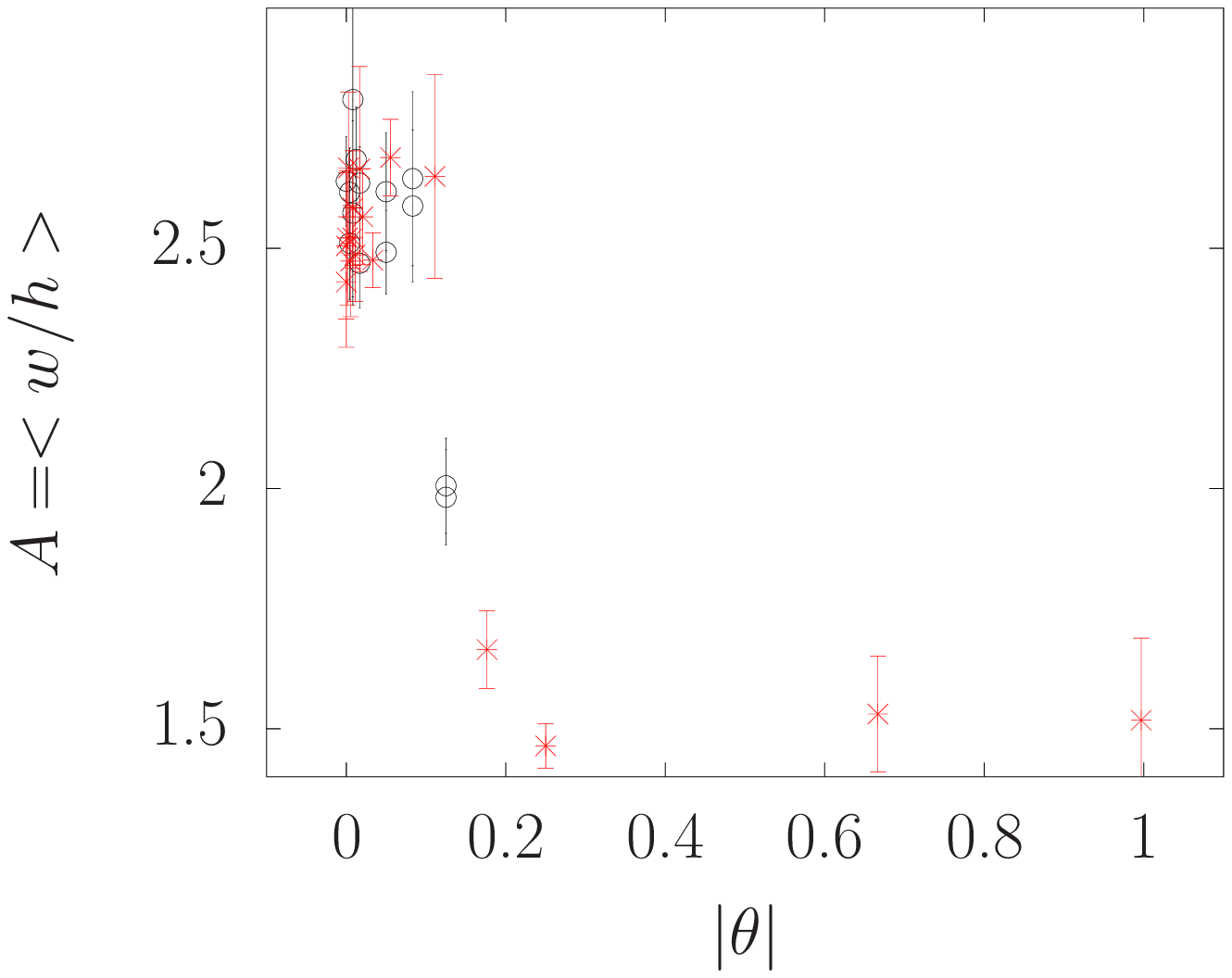}\\
\end{tabular}
\caption{(Color online) Left, experimental spatio-temporal
spectrum $E_E(k_x,\theta)$ at $\mathrm{Re}\approx 10^5$. The shear
layer reaches the slower impeller at $\theta =\pm 0.1$. Colors are
log scaled. Right, $A=<w/h>$ as defined in
Fig.\,\ref{fig:isotropy} as a function of $|\theta|$ at
$\mathrm{Re}\approx 10^5$ (black circles) and $\mathrm{Re}\approx
5\times 10^5$ (red/gray stars).} \label{fig:isotropy2}
\end{figure}

It is tempting to associate the Eckhaus azimuthal instability
evidenced in the previous section to a Kelvin-Helmoltz instability
of the shear layer. Indeed, this instability was already pointed
out as the physical mechanism for a stability loss of the
axi-symmetric steady-state followed by successive flows with
non-zero azimuthal wavenumber. This was numerically observed for
small Reynolds number ($\mathrm{Re}<500$) in a
geometry\,\cite{nore_12_2003} close to ours. Here we report a
similar mechanism but at largely higher Reynolds numbers. It was
also observed that the Kelvin-Helmoltz instability results in a
series of large scale structures taking the shape of radial
vortices with well defined azimuthal wavenumber that can be seen
e.g. through bubble seeding and white
light.\cite{ravelet_supercritical_2008,cortet_normalized_2009} The
possibility of changes in $m$ of this pattern revealing an Eckhaus
instability analog to that existing in small dynamical systems was
already mentioned in Ref.~\onlinecite{cortet_susceptibility_2011}.
To sustain this possibility, we have monitored the 2D-spatial
spectrum at a fixed Reynolds number, as a function of $\theta$. It
was already reported (see for example
Ref.~\onlinecite{cortet_susceptibility_2011}) a shifting of the
shear layer toward the slower impeller with increasing $|\theta|$.
When $|\theta| > 0.1$ the shear layer was found to be completely
absorbed. If the azimuthal structure we observed are related to
the shear layer, the structure of the 2D-spatial spectrum should
also follow $\theta$. This is indeed the case, as can be seen in
Fig.\,\ref{fig:isotropy2} both from the isotropy  and from the
1D-spatial spectrum $E_E(k_x,\theta)$. With increasing $|\theta|$
the two bands of $k_x$ merge into one, at $k_x=0$. At $|\theta|=
0.1$ the isotropy measurement and the peak at $k_x=0$ of the
spectrum drops suddenly. Further increasing $\theta$, both the
isotropy and the spatial spectrum (not shown) are found to be
independent of $\theta$. This is indicative of a transition into a
$m=0$ structure, with no azimuthal modulations anymore. This thus
proves that the azimuthal modulation is indeed strongly associated
with the shear layer.

\subsection{Visualization of the coherent structure}
\label{subsec:vizu}
\begin{figure}[h]
\centerline{\includegraphics[width=0.345\textwidth]{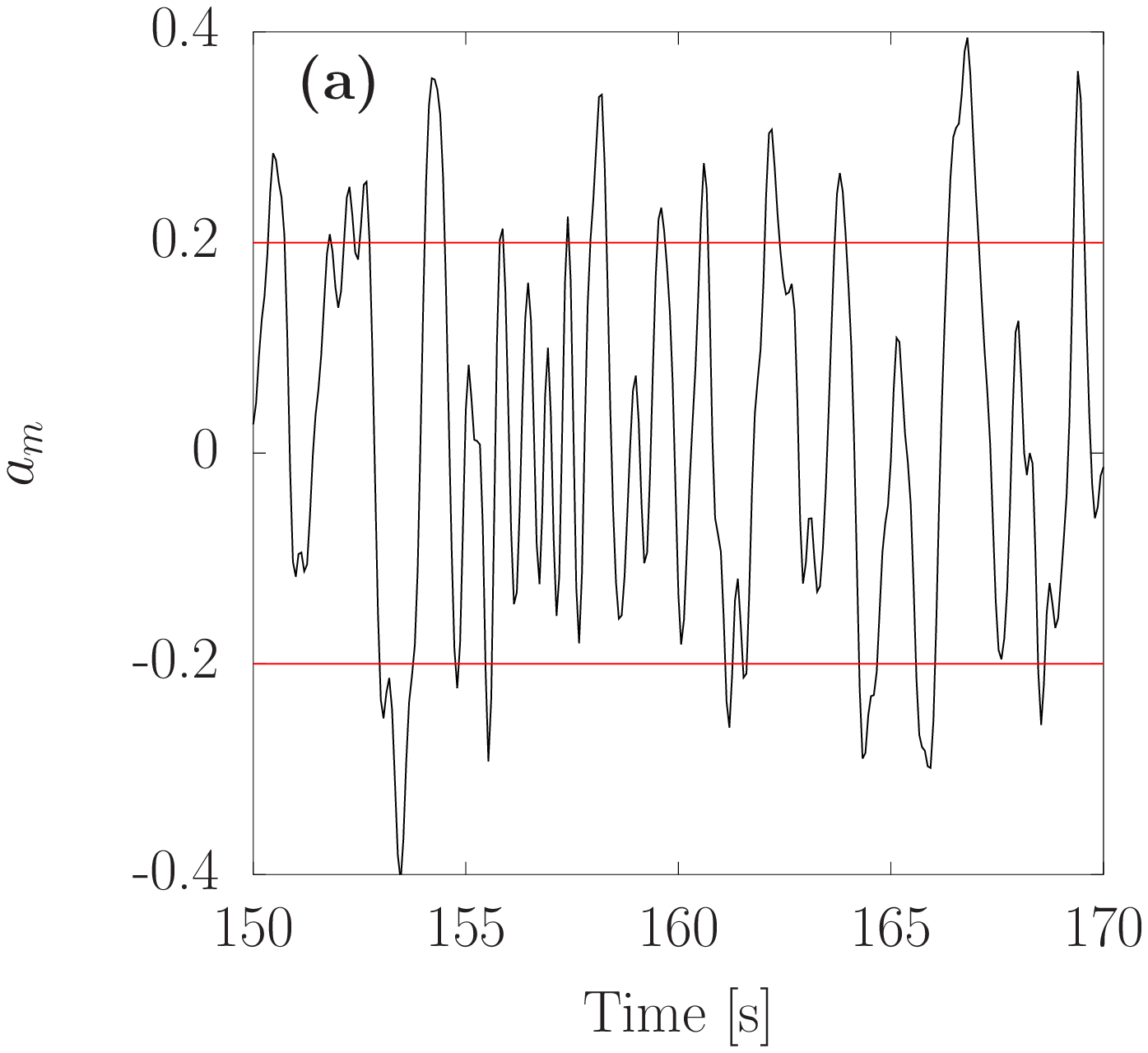}\hspace{1cm}\includegraphics[width=0.34\textwidth]{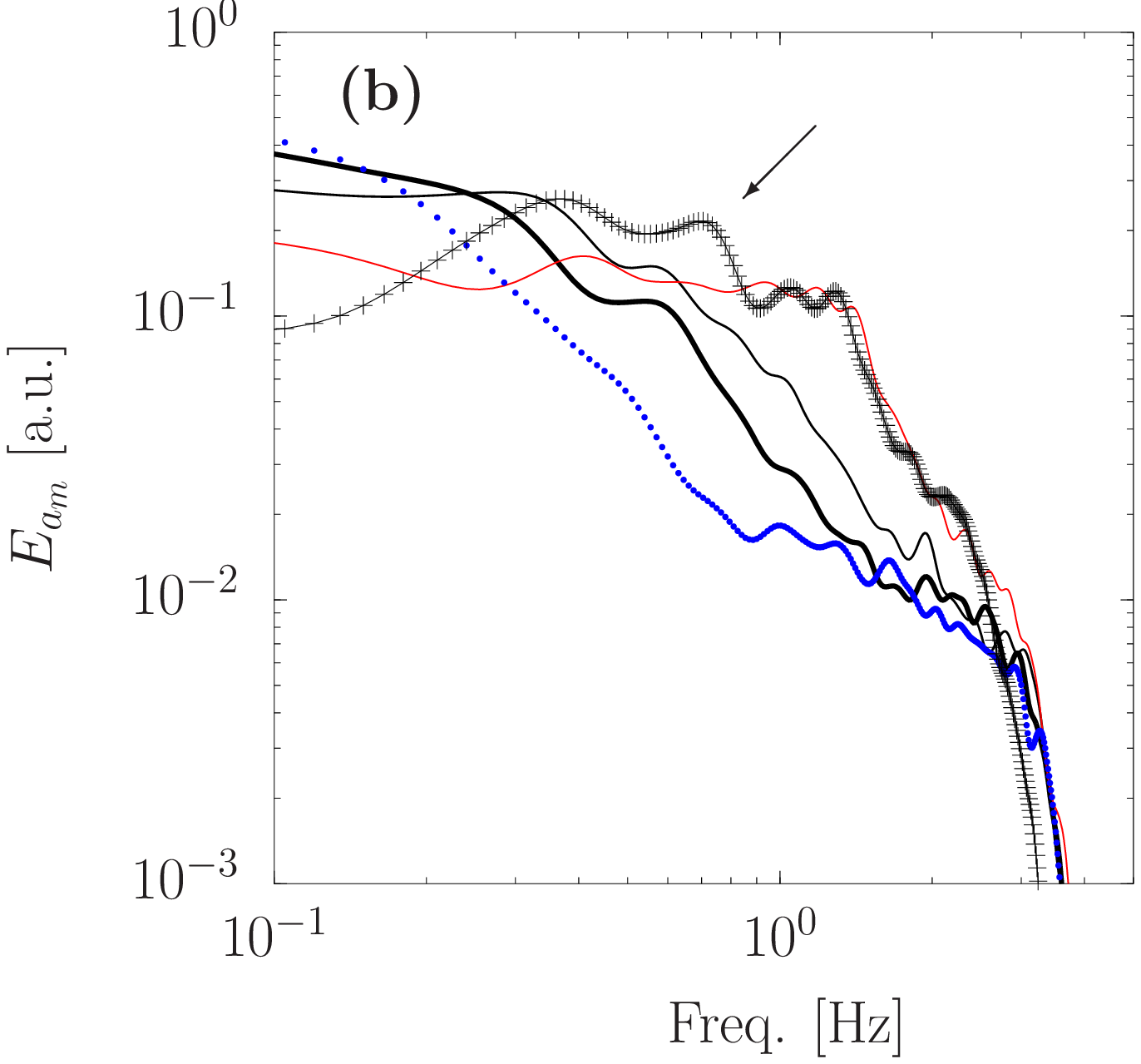}}
\caption{(Color online) (a) $a_m(t)=\int xdx\int dz {\bf v'}\cdot
{\bf \mv^{1mk_0}} $ with $m=3$ and $\mathrm{Re}=10^6$  at a fixed
angle $\varphi$ (corresponding to the angle of
Fig.\,\ref{fig:velocityfield}(c)) as a function of time. $a_m(t)$
was computed using ${\bf v'}(t)$ from the time-series used in
Fig.\,\ref{fig:azimuthalModes}(d) and was finally low-pass
filtered at 4\,Hz. Horizontal lines, see
Fig.\,\ref{coherentstructure}. (b) power spectrum $E_{a_m}(f)$ of
$a_m(t)$ ($+$ solid line) for the data shown in (a) and for other
Reynolds numbers: $\mathrm{Re}/1000 = 41$ (thin red/gray line), 21
(thin black line), 12 (thick black line), 7 ($\circ$). The arrow
marks the main peak evolution with decreasing Re.} \label{fig:am}
\end{figure}

Finally the existence of organized motions in
the turbulent flow means ${\bf v'}$ can be written as the sum of
an in-phase term and a fluctuation term: ${\bf v'}={\bf v_c}+{\bf
v''}$, with  ${\bf v''}$ the fluctuations and ${\bf v_c}$ the
coherent structure velocity field. A visualization of the spatial
structure of one azimuthal mode $m$ can then be obtained using
conditional average of the velocity field ${\bf v'}$. The
condition is given by suitable projection onto the corresponding
Beltrami mode, see Eq.\,(\ref{projection}). Due to the geometry of
our PIV device, we however only have access to the velocity field
in a meridional plane, i.e. for $\varphi_0=[0,\pi]$. Since the
azimuthal modes are rotating we may however replace the azimuthal
integration in the scalar product by a suitable time integration.
This is made possible for high enough Reynolds number where the
rotation frequency of the coherent structure is resolved. We used
conditional average as follows: we first compute for each
instantaneous experimental field ${\bf v'}$ the quantity:
\begin{equation}
a_{m}(t)
=\int_{-1}^1 x\,dx \int_{-h/R}^{h/R} dz \, {\bf v'}(t)\cdot 
{\bf\mv^{1mk_0}}(x,\varphi_0,z), \label{semiprojection}
\end{equation}
keeping the azimuthal wavenumber $m$ and the angle $\varphi$
constant. A result is shown in Fig.\,\ref{fig:am}(a) obtained
using the same experimental data shown in
Fig.\,\ref{fig:azimuthalModes}(d) at $\mathrm{Re}=10^6$ and the
synthetic velocity field as in Fig.\,\ref{fig:velocityfield}(d) at
$m=3$. One sees quasi-sinusoidal fluctuations in the signal
$a_m(t)$. We propose that these fluctuations correspond to the
coherent structures passing through the PIV plane. This is
corroborated by looking at $E_{a_m}(f)$, the temporal power
spectrum of $a_m(t)$ in Fig.\,\ref{fig:am}(b). The rotation
frequency measured using the 2D-spectrum of the velocity field and
the  main peak frequency of $E_{a_m}$ ($f=0.73$\,Hz in this
experiment) are found identical. Secondary peak is a sub-harmonic
at $f=0.37$\,Hz. For lower Reynolds number, below
$\mathrm{Re}=4\times 10^4$, both amplitudes of these peaks
decrease and are shifted toward lower frequencies. Simultaneously
the energy becomes roughly equally distributed along a plateau,
which amplitude is increasing and width is decreasing with the
Reynolds number. Down to $\mathrm{Re}= 10^4$ a peak at small
amplitude  can  still be distinguished at $f\approx 0.5$\,Hz.  For
the lower Reynolds number, the plateau vanishes and the energy is
mainly concentrated in the vicinity of $f=0$. This confirms the
behavior described in section\,\ref{subsec:tempbehav}.

\begin{figure}[h]
\centerline{\includegraphics[width=0.35\textwidth]{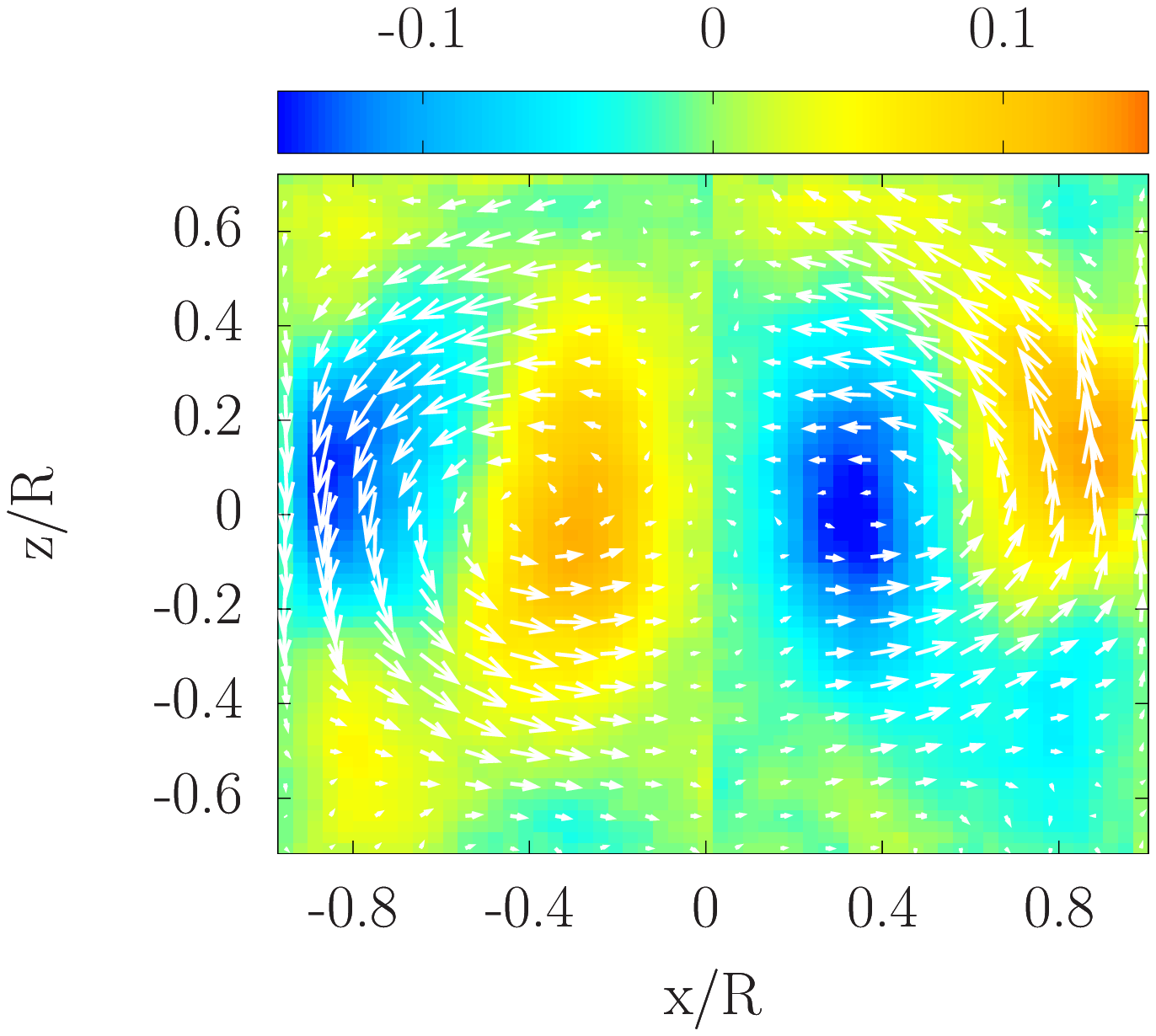}
\hspace{1cm}\includegraphics[width=0.35\textwidth]{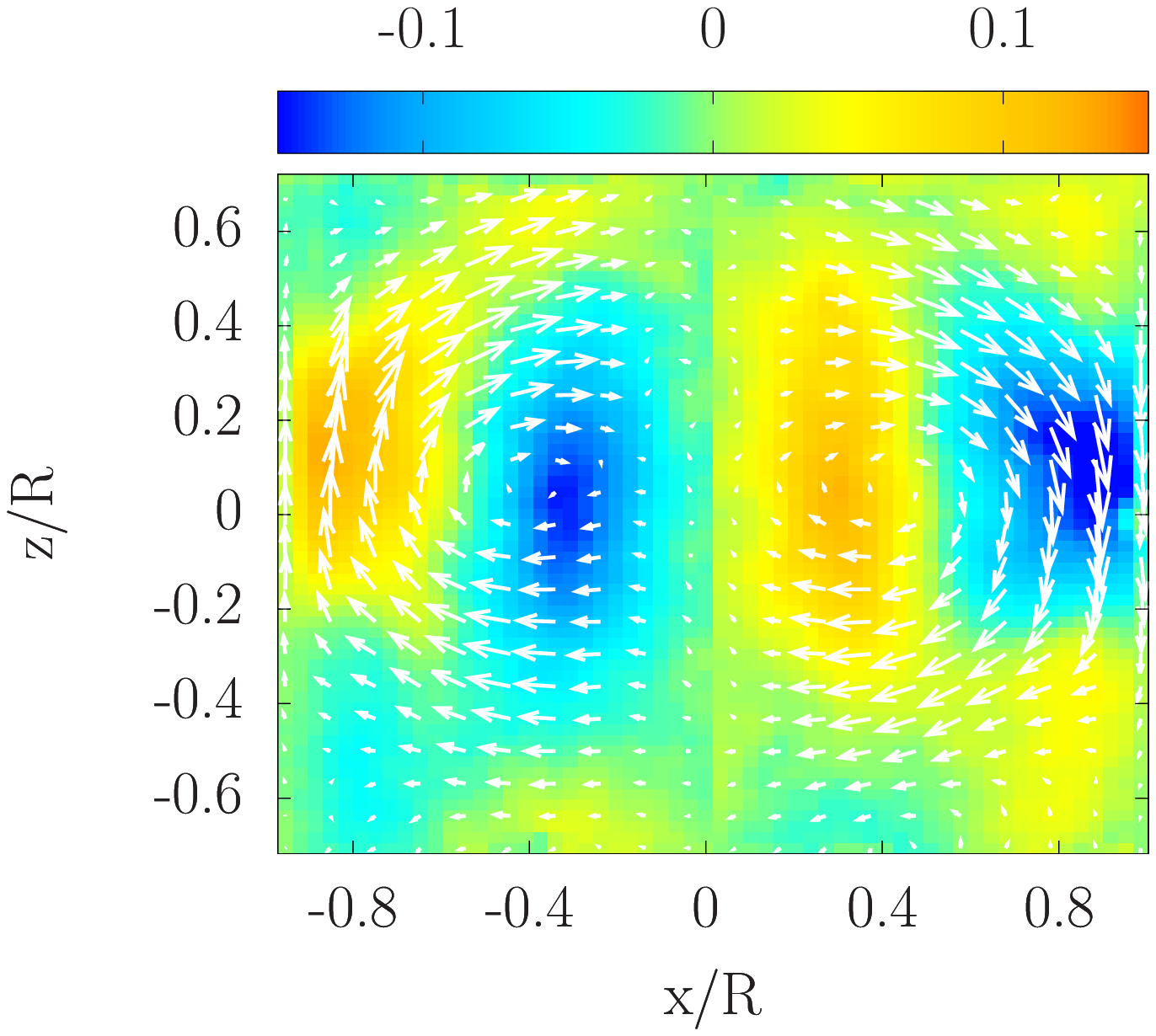}}
\caption{ (Color online) Velocity fields obtained with the
conditional average (left) ${\bf v'}(t[a_m(t)>\epsilon])$ in the
plane $\varphi=[0;\pi]$ and (right) ${\bf v'}(t[a_m(t)<-\epsilon])$ in
the orthogonal plane $\varphi=[\pi/2;3\pi/2]$. Both are computed using the
signal $a_m(t)$ and the horizontal lines represented in
Fig.\,\ref{fig:am}(a) for $\mathrm{Re}=10^6$, $m=3$ and
$\epsilon=0.2$. We clearly distinguish two pairs of azimuthal
vortices, rotating around the azimuthal axis. The observed discontinuity at $x=0$ is due to a slight shift of our S-PIV laser sheet from the center of the tank. (enhanced online)}
\label{coherentstructure}
\end{figure}

The velocity field ${\bf v'}$ was then averaged over times for
which $a_m$ lies in a given range of values $[a_m>\epsilon]$,
representative of a given phase of the velocity field. An example
is provided in Fig.\,\ref{coherentstructure} for $\epsilon=0.2$
and $m=3$ from $a_m(t)$ shown in Fig.\,\ref{fig:am}. One sees the
presumable $m=3$ coherent structure in two planes, dephased by
$\pi$.  One indeed sees large scale structures looking like pairs
of azimuthal vortices. 
By reproducing this conditional average
using different phase $\varphi_0$ for $\mv^{1mk_0}$ at a fixed
$m$, we obtained the complete movie (see Fig.\,\ref{coherentstructure}) of the rotation
of this pair of vortices (enhanced online). 
The physical mechanism leading to this
vortices can be the Kelvin-Helmoltz instability laying in the
shear layer, a secondary or related instability or the
 centrifugal instability. It is interesting to compare this
result with the $m=3$ velocity field shown in
Fig.\,\ref{fig:velocityfield}(d) using the same phase $\varphi_0$
as in Fig.\,\ref{coherentstructure}(a). Both the anti-symmetric
nature of the velocity field with respect to the $r=0$ vertical
axis and ${\bf v}(r=0) = 0$ are recovered.

\section{Some theoretical consequences}
\label{sec:theocons} In 1970, Kraichnan wrote a paper entitled
``Instability in fully developed
turbulence",~\cite{kraichnan_instability_1970} in which he focuses
on the propagation of errors from one range of scale to another.
This problem is related to predictability of turbulence. In the
present work, we consider another aspect of the same problem
related to the stability of the unsteady coherent structures of a
turbulent flow (in other words, to the dynamic of the perturbation
of this given coherent structures). We have shown that using
spatio-temporal spectra, and comparison or projection onto
suitable Beltrami modes, it is possible to identify coherent
perturbations to the time-averaged flow. This is liable to an
Eckhaus instability analog to what is observed in systems in the
laminar limit, with a few degrees of freedom. In classical
phenomenology of turbulence the number of degrees of freedom
scales like $\mathrm{Re}^{9/4}$. Given our range of Reynolds
number (from $10^2$ to $10^6$), we thus can expect our system to
be described by a number of degrees of freedom in the range
$7\times 10^4$ to $3\times 10^{13}$. Our results thus prove that a
huge  number of degrees of freedom are irrelevant to describe the
instability of our turbulent flow. This sets a number of
interesting questions, that we list below.

In the present paper, we have characterized experimentally the
instability of the basic state. Is there any analytical or
numerical way to explain our observations? Classical instability
analysis for a given velocity field ${\bf v}(x,t)$ starts from
stationary solutions of the Navier-Stokes equation:
\begin{eqnarray}
\partial_j v_j&=&0,\nonumber\\
\partial_t v_i+v_j\partial_j v_i&=&-\frac{1}{\rho}\partial_i p 
+f_i+\nu\partial_k\partial_k
v_i.
\label{NSE}
\end{eqnarray}
In our case, we consider the instability of a time-average basic
state $\bar {\bf v}(x,t)$, that obeys the equation:
\begin{eqnarray}
\partial_j \bar v_j&=&0,\\
\partial_t \bar 
v+\bar{v_j}\partial_j\bar{v_i}&=&-\frac{1}{\rho}\partial_i
\bar p +\bar f_i+\nu\partial_k\partial_k \bar v_i +\partial_j
R_{ij}, \nonumber
\label{NSEcg1}
\end{eqnarray}
where $\rho\,R_{ij}=\rho\,(\bar v_i \bar v_j-\overline{v_i v_j})$
is the Reynolds stress tensor.  This Reynolds stress represents the
influence of all the degrees of freedom of the flow onto its
average, and can, in general, only be computed via full solution
of the NS equation. Therefore, the problem of instability of a
mean turbulent flow cannot be tackled analytically or is too
demanding numerically, unless a prescription (parametrization) of
the Reynolds stress is provided. In the case of the plane Couette
turbulent flow, for example, this was attempted by Tuckerman
\emph{et al.}~\cite{tuckerman_instability_2010} via the K -
$\Omega$ closure model. They calculate steady 1D solution profiles
of the K - $\Omega$ model and their linear stability to 3D
perturbations, but find no correspondence between this analysis
and the onset of turbulent-laminar bands in experiments and
simulations. In the same way, Legras and
Villone\,\cite{legras_stability_2005} address the problem of
Kolmogorov flow instability when molecular viscosity is replaced
by a Smagorinsky parametrization for small-scale turbulence. Such
a parametrization represents the motion at scales smaller than the
large scale Kolmogorov flow. They claim that it may  provide hints
on large-scale instabilities at large $\mathrm{Re}$ and,
hopefully, on the character of such instabilities, but it has
never been checked. In our case, it was observed that the sequence
of Eckhaus instability resembles the sequence of instability
observed at much lower Reynolds number.~\cite{nore_12_2003} This
suggest that the replacement of the Reynolds stress by an
eddy-viscosity might be a suitable parametrization by increasing
the dissipation, and therefore decreasing the effective Reynolds
number (and the effective number of degrees of freedom).
Theoretically, we note that the special geometry of our experiment
and its parity symmetry properties make it plausible that the
first term in the expansion of the Reynolds stress as a function
of the velocity gradient is indeed quadratic, resulting in a
non-isotropic eddy-viscosity tensor (in other words, there is
probably no anisotropic kinetic alpha (AKA) term in the Reynolds
stress expansion).~\cite{dubrulle_eddy_1991} It is however not
clear that the correct parameters of the instability (threshold,
wavenumbers) can be captured with a simple eddy-viscosity model,
since this procedure fails in the case of a plane Couette
flow.~\cite{tuckerman_instability_2010} An interesting alternative
would be to derive directly the mean state as the critical points
of a suitable Arnold functional, making the problem liable to
traditional tolls of bifurcation theory. Such strategy was
followed in Ref.~\onlinecite{briceNJP}, resulting in a full
description of a spontaneous parity-breaking turbulent
bifurcation. The generalization of this model to non-axisymmetric
mean state is however an open problem.

Summarizing, we have shown that the fluctuations of a turbulent
von K\'arm\'an flow obey a sequence of Eckhaus-like instabilities with
varying Reynolds number that is similar to the sequence of Echaus
instability observed in the laminar flow, at much lower Reynolds
number and that may be explained through a suitable
parametrization by increasing the dissipation, and therefore
decreasing the effective Reynolds number (and the effective number
of degrees of freedom). If this is indeed true, it may have
interesting implications for other fields, especially astrophysics
and geophysics where Reynolds number are huge and our interest is
mainly in the dynamics of the large-scale mean flow. For example,
current climate models that have presently very low effective
Reynolds number may be able to capture efficiently certain types
of instabilities of the mean (climatic) turbulent state. They may
therefore be more predictable that Kraichnan thought, in a very
different meaning.

\acknowledgments E. Herbert acknowledges the support of the
program DSM-Energie and is grateful to  V.\,Padilla and
C.\,Wiertel-Gasquet for technics and B.\,Saint-Michel for
interesting discussions.


\bibliographystyle{abbrv}

\end{document}